\documentclass[12pt]{article}
\usepackage{amsfonts}
\usepackage{amsmath}
\usepackage{amssymb}
\usepackage{array}
\usepackage{bigints}
\usepackage{booktabs}
\usepackage[nosort]{cite}
\usepackage{color}
\usepackage{dsfont}
\usepackage{float}
\usepackage{framed}
\usepackage{graphicx}
\usepackage{indentfirst}
\usepackage{mathrsfs}
\usepackage{multirow}
\usepackage{setspace}
\usepackage{subdepth}
\usepackage{subfig}
\usepackage{titlesec}
\usepackage[dotinlabels]{titletoc}
\usepackage{wrapfig}
\usepackage[all]{xy}
\usepackage{young}
\usepackage[vcentermath]{youngtab}

\usepackage{hyperref}
\hypersetup{colorlinks=true}
\hypersetup{linkcolor=black}
\hypersetup{citecolor=black}
\hypersetup{urlcolor=black}

\numberwithin{equation}{section}

\usepackage{tikz}
\usetikzlibrary{arrows,shapes,positioning}
\usetikzlibrary{decorations.markings}
\usepackage[rightcaption]{sidecap}
\tikzstyle arrowstyle=[scale=1]
\tikzstyle directed=[postaction={decorate,decoration={markings,
    mark=at position .65 with {\arrow[arrowstyle]{stealth}}}}]
\tikzstyle reverse directed=[postaction={decorate,decoration={markings,
    mark=at position .65 with {\arrowreversed[arrowstyle]{stealth};}}}]
\usetikzlibrary{positioning}

\usepackage[left=2.5cm,right=2.5cm,top=2.5cm,bottom=3cm]{geometry}
\titleformat{\section}{\normalfont\bfseries}{\thesection.}{4pt}{}
\titlespacing{\section}{0pt}{20pt}{6pt}

\titleformat{\subsection}{\normalfont\itshape}{\thesubsection.}{4pt}{}
\titlespacing{\subsection}{0pt}{15pt}{6pt}

\titleformat{\subsubsection}{\normalfont}{\thesubsubsection.}{4pt}{}
\titlespacing{\subsubsection}{0pt}{15pt}{6pt}

\def\tilde{\widetilde}
\def\t{\tilde}
\def\hat{\widehat}

\def\half{{1 \over 2}}
\def\d{\partial}

\def\ep{\varepsilon}
\def\1{{\mathds 1}}

\DeclareMathOperator{\tr}{tr}

\newcommand{\Z}{{\mathbb Z}}
\newcommand{\C}{{\mathbb C}}
\newcommand{\R}{{\mathbb R}}

\def\CF{{\mathcal F}}

\def\CI{{\mathcal I}}

\def\CN{{\mathcal N}}
\def\CO{{\mathcal O}}

\DeclareFontShape{OT1}{cmr}{mx}{n}%
    {<->cmr10}{}
\newcommand{\mytitlefont}{\fontseries{mx}\selectfont}
\DeclareMathAlphabet{\titlemath}{OT1}{cmr}{mx}{n}

\begin{document}


\begin{titlepage}

\begin{center}

~\\[2cm]

{\fontsize{22pt}{0pt} \mytitlefont Revisiting the Multi-Monopole Point of  \\[5pt]  
$SU(N)$~$\CN=2$ Gauge Theory in Four Dimensions}

~\\[15pt]

Eric D'Hoker, Thomas T.~Dumitrescu, Efrat Gerchkovitz, and Emily Nardoni

\bigskip

{\it Mani L.\,Bhaumik Institute for Theoretical Physics,}\\[-2pt]
{\it Department of Physics and Astronomy,}\\[-2pt]
       {\it University of California, Los Angeles, CA 90095, USA}

~\\[15pt]

\end{center}

\noindent Motivated by applications to soft supersymmetry breaking, we revisit the expansion of the Seiberg-Witten solution around the multi-monopole point on the Coulomb branch of pure~$SU(N)$~$\CN=2$ gauge theory in four dimensions. At this point~$N-1$ mutually local magnetic monopoles become massless simultaneously, and in a suitable duality frame the gauge couplings logarithmically run to zero. We explicitly calculate the leading threshold corrections to this logarithmic running from the Seiberg-Witten solution by adapting a method previously introduced by D'Hoker and Phong. We compare our computation to existing results in the literature; this includes results specific to~$SU(2)$ and~$SU(3)$ gauge theories, the large-$N$ results of Douglas and Shenker, as well as results obtained by appealing to integrable systems or topological strings. We find broad agreement, while also clarifying some lingering inconsistencies. Finally, we explicitly extend the results of Douglas and Shenker to finite~$N$, finding exact agreement with our first calculation.

\vfill

\begin{flushleft}
December 2020
\end{flushleft}

\end{titlepage}


\tableofcontents

\newpage

\section{Introduction}

\subsection{The Multi-Monopole Point of $SU(N)$ $\CN=2$ Gauge Theory}\label{ssec:mmp}

Since the work of Seiberg and Witten~\cite{Seiberg:1994rs,Seiberg:1994aj}, non-perturbative~$\CN=2$ gauge dynamics has been a topic of central importance in quantum field theory (QFT), with deep connections to string theory and mathematics. In~\cite{Seiberg:1994rs} the authors solved for the low-energy effective QFT on the Coulomb branch of pure $SU(2)$~$\CN=2$ gauge theory in four dimensions. At generic points on the Coulomb branch, this low-energy theory is described by a single~$U(1)$~$\CN=2$ vector multiplet, whose leading interactions are encoded by its complexified gauge coupling~$\tau(u)$. Here~$u \sim \tr \phi^2$ is a gauge-invariant coordinate on the Coulomb branch, with~$\phi$ the complex~$SU(2)$ adjoint Lorentz scalar residing in the~$\CN=2$ vector multiplet. Crucially, $\tau(u)$ may undergo~$SL(2, \Z)$ electric-magnetic duality transformations as~$u$ traverses closed loops in the~$u$-plane. 

The function~$\tau(u)$ was constructed by identifying it with the modular parameter of an auxiliary, $u$-dependent Riemann surface~$\Sigma$ of genus one -- the Seiberg-Witten curve. This function is closely related to the special Coulomb-branch coordinates~$a(u), a_D(u)$, which are determined by period integrals of a suitable meromorphic one-form (the Seiberg-Witten differential) along canonical~$A$- and~$B$-cycles of~$\Sigma$. Once the special coordinates are known, the gauge coupling can be computed via~$\tau =- {d a_D \over d a}$.\footnote{{~We use conventions in which~$\tau= - \frac{da_D}{da}$ and~$\tau_D = -{1 \over \tau} = \frac{da}{da_D}$. This differs by an overall sign from the more familiar conventions (e.g.~used in~\cite{Seiberg:1994rs}) in which~$\tau = {da_D \over da}$ and~$\tau_D = - {1 \over \tau} = - {da \over da_D}$. This difference arises because our~$a$-periods differ from those in~\cite{Seiberg:1994rs} by a minus sign, while our~$a_D$-periods agree.}} The choice of canonical~$A$- and~$B$-cycles is arbitrary, and different choices are related by~$SL(2, \Z)$ duality transformations of the special coordinates and~$\tau$. The special coordinates also determine the masses of heavy BPS particles. A BPS particle with electric and magnetic charges~$(q_e, q_m) \in \Z$ has mass~$M_\text{BPS} \sim |q_e a + q_m a_D|$. Note that the electric special coordinate~$a$ is the scalar residing in the low-energy~$U(1)$ $\CN=2$ vector multiplet. 

An important feature of the~$SU(2)$ Seiberg-Witten solution~\cite{Seiberg:1994rs} is that the curve~$\Sigma$ degenerates at two points~$u \sim  \pm \Lambda^2$ of the~$u$-plane. These two points are related by a discrete~$\Z_8$~$R$-symmetry, which maps~$u \rightarrow -u$. Here~$\Lambda$ is the strong-coupling scale of the~$SU(2)$ $\CN=2$ gauge theory. At these points the gauge coupling diverges and there are additional massless particles: a magnetic monopole with~$(q_e, q_m) =(0,1)$ at $u \sim \Lambda^2$, and a dyon with~$(q_e, q_m) =(2,1)$ at $u \sim -\Lambda^2$. These points are, respectively, known as the monopole and dyon points of the~$SU(2)$ theory. Since these points are exchanged by the spontaneously broken~$\Z_8$~$R$-symmetry, the low-energy physics at the two points is the same. As is customary, we will focus on the monopole point. Near this point, this theory is most conveniently described in terms of~$S$-dual magnetic variables: a~$U(1)_D$ $\CN=2$ vector multiplet, with scalar component~$a_D$ and gauge coupling
\begin{equation}
\tau_D = {d a \over d a_D}~.
\end{equation}
The unit monopole is a BPS state of mass~$M_\text{BPS} \sim a_D$, so that the monopole point is given by~$a_D = 0$. There the monopole can be described by coupling the~$U(1)_D$ vector multiplet to a massless hypermultiplet carrying unit electric charge under~$U(1)_D$. This renders the dual magnetic gauge coupling IR free and drives it to zero logarithmically, which implies the following behavior for~$\tau_D$ near the monopole point, 
\begin{equation}
\label{eq:tdirlog}
\tau_D (a_D) = -{i \over 2 \pi} \log a_D + (\text{regular}) \quad \text{as} \quad a_D \rightarrow 0~.
\end{equation}
The coefficient of the logarithm is fixed by the unit charge of the massless monopole, while its branch cut ensures the correct~$SL(2, \Z)$ monodromy around the monopole point. The same phenomenon occurs at the dyon point, except that the simple IR free description occurs in a different duality frame. 

The monopole and dyon points of the~$SU(2)$ $\CN=2$ theory play a crucial role in many applications of Seiberg-Witten theory. For instance, it was shown in~\cite{Seiberg:1994rs} that they describe the two confining vacua of the pure~$SU(2)$ $\CN=1$ gauge theory obtained by adding the $\CN=2 \rightarrow \CN=1$ breaking superpotential~$\int d^2 \theta \, u \sim \int d^2 \theta \, \tr \phi^2 $ via Higgsing in the IR free~$U(1)_D$ gauge theory described above. In applications of~$\CN=2$ gauge theory to four-manifold topology, the monopole and dyon points give rise to the Seiberg-Witten equations~\cite{Witten:1994cg}.

In this paper we are interested in the generalization of the~$SU(2)$ monopole and dyon points to pure~$SU(N)$ $\CN=2$ gauge theories.  A systematic study of these points was initiated in~\cite{Douglas:1995nw}, building on the~$SU(N)$ generalization of the Seiberg-Witten solution found in \cite{Klemm:1994qs,Argyres:1994xh,Klemm:1994qj,Klemm:1995wp}. The Coulomb branch is now~$N-1$ complex dimensional and described by the gauge-invariant coordinates~$u_n \sim \tr\phi^n~(n = 2, \ldots, N)$, collectively denoted by~$u$. (As before, $\phi$ is the complex~$SU(N)$ adjoint and Lorentz scalar in the~$\CN=2$ vector multiplet.) The low-energy effective theory at generic points is a~$U(1)^{N-1}$ gauge theory, and there are~$N-1$ dual pairs of special coordinates~$a_k(u), a_{Dk}(u)~(k =1, \ldots, N-1)$. They are the~$A$- and~$B$-cycle periods of a suitable meromorphic differential~$\lambda$ on the~$u$-dependent Seiberg-Witten curve~$\Sigma$, which now has genus~$N-1$. As before, the special coordinates determine the matrix~$\tau_{k\ell}$ of complexified~$U(1)^{N-1}$ gauge couplings via~$\tau_{k\ell} =- {\d a_{Dk} \over \d a_{\ell}}$, and the masses of BPS states with charges~$(q_{e k}, q_{m\ell}) \in \Z^{2(N-1)}$ via~$M_\text{BPS} \sim \big|\sum_{k=1}^{N-1} \left(q_{e k} a_k + q_{m k} a_{Dk}\right)\big|$. Changing the choice of canonical~$A$- and~$B$-cycles on~$\Sigma$ acts on the special coordinates and the matrix of couplings via an~$Sp(2N - 2, \Z)$ electric-magnetic duality transformation.

As was explained in~\cite{Douglas:1995nw}, the Coulomb branch of the~$SU(N)$ gauge theory has many interesting singular points, at which the Seiberg-Witten curve~$\Sigma$ degenerates in various ways. The BPS dyons that become massless at such points are typically mutually non-local, i.e.~they have non-vanishing Dirac pairing~$\sum_{k = 1}^{N-1} \left(q_{ek} q'_{mk} - q'_{ek} q_{mk}\right)$. In particular, this means that there is no electric-magnetic duality frame in which all of them carry electric charges. Such mutually non-local massless dyons describe interacting superconformal field theories~\cite{Argyres:1995jj,Argyres:1995xn}.

By contrast, the singular points that generalize the monopole and dyon points of the~$SU(2)$ theory arise when~$N-1$ (i.e.~the maximal number of) mutually local BPS dyons simultaneously become massless~\cite{Douglas:1995nw}. This happens at precisely~$N$ distinct points on the Coulomb branch, which are related by a spontaneously broken~$\Z_{4 N}$ $R$-symmetry, which rotates the Coulomb branch coordinates~$u_n$ by~$N$-th roots of unity. We will collectively refer to these~$N$ points on the Coulomb branch as the multi-dyon points of the~$SU(N)$ theory. As before, it is sufficient to focus on one such point, and we choose the multi-monopole point. At the multi-monopole point the~$N-1$ mutually local massless dyons are electrically neutral and carry unit magnetic charge in precisely one~$U(1)$ factor of the low-energy gauge group. 

As in the~$SU(2)$ theory, it is useful to pass to an~$S$-dual magnetic description, which is a~$U(1)_D^{N-1}$ $\CN=2$ gauge theory with vector-multiplet scalars~$a_{D k}$ and gauge coupling matrix
\begin{equation}\label{eq:tdaadmat}
\tau_{D k\ell} = {\d a_k \over \d a_{D \ell}}~, \qquad (k, \ell = 1, \ldots, N-1)~.
\end{equation}
The~$k$-th unit monopole is a BPS state of mass~$M_\text{BPS} \sim a_{Dk}$, and hence the multi-monopole point is given by~$a_{Dk} = 0$ for all~$k$. There each monopole is described by a massless hypermultiplet that carries unit electric charge under the~$k$-th~$U(1)_D$ gauge factor, and is neutral with respect to the~$N-2$ other~$U(1)_D$ factors. As in~\eqref{eq:tdirlog}, this completely determines the singular behavior of~$\tau_{D k \ell}$ near the multi-monopole point,
\begin{equation}\label{eq:tdmatirlog}
\tau_{D k\ell} = -{i \over 2 \pi} \delta_{k\ell} \log a_{Dk} + (\text{regular}) \quad \text{as} \quad a_{D k} \rightarrow 0~.
\end{equation}

As before, the addition of the~$\CN=2 \rightarrow \CN=1$ preserving superpotential~$\int d^2 \theta \, u_2  \sim \int d^2 \theta \, \tr \phi^2$ collapses the Coulomb branch of the~$\CN=2$ theory to the~$N$ multi-dyon points, correctly capturing the~$N$ vacua of the pure~$SU(N)$ $\CN=1$ gauge theory~\cite{Douglas:1995nw}. Moreover, the~$N-1$ massless monopoles Higgs the~$U(1)_D^{N-1}$ gauge theory, leading to confinement. These conclusions do not depend on the structure of the regular terms in~\eqref{eq:tdmatirlog}. They only rely on the massless matter content of the~$U(1)_D^{N-1}$ gauge theory at the multi-monopole point (which is reflected in the logarithmic terms in~\eqref{eq:tdmatirlog}), as well as on the fact that the~$a_k$ special coordinates at the multi-monopole point are all non-zero.\footnote{~To see this, recall from~\cite{Seiberg:1994aj,Douglas:1995nw} that the monopole vev responsible for Higgsing the~$k$-th~$U(1)_D$ factor of the gauge group is set by~${\d u_2 \over \d a_{D k}}$ at the multi-monopole point~$a_{D k} = 0$. To evaluate this, it is convenient to use the renormalization group equation~$u_2(a_D) \sim (\sum_{k = 1}^{N-1} a_k a_{Dk} - 2\CF_D(a_D))$ derived in   \cite{Matone:1995rx,Eguchi:1995jh, Bonelli:1996qc, Bonelli:1996qh, Howe:1996pw,DHoker:1996yyu}. (See~\cite{Luty:1999qc} for a simple derivation that involves promoting~$\Lambda$ to an~$\CN=2$ chiral background superfield.)  Here~$\CF_D$ is the dual prepotential, so that~${\d \CF_D \over \d a_{D k}} = a_k$. Using~\eqref{eq:tdaadmat}, we then find that
\begin{equation}
{\d u_2 \over \d a_{D k}} \sim {\sum _{\ell =1}^{N-1}} \tau_{D k\ell} a_{D \ell} - a_k~.
\end{equation}
It follows from~\eqref{eq:tdmatirlog} that the first term vanishes at the multi-monopole point, leaving only the term~$\sim a_k$.} These were first computed in~\cite{Douglas:1995nw}, 
\begin{equation}\label{eq:akatmmpi}
a_k(a_{D \ell} = 0) \sim N \Lambda \sin{k \pi \over N}~,
\end{equation} 
and they are indeed all non-vanishing. We will recover this result below, including a scheme-dependent prefactor that we omit here.\footnote{~Rescaling $\Lambda$ by a constant amounts to a change of renormalization scheme.} We now turn to applications of Seiberg-Witten theory that are sensitive to the regular terms in~\eqref{eq:tdmatirlog}. 

\subsection{Motivation and Summary of Results}

The computations described in this paper were motivated by applications of Seiberg-Witten theory that require more detailed information about the multi-monopole point than the leading logarithmic running of the couplings in~\eqref{eq:tdmatirlog} or the value of the~$a_k$-periods in~\eqref{eq:akatmmpi}. (Two such applications are mentioned below.) Our primary interest will be the leading regular terms in~\eqref{eq:tdmatirlog}, which we parametrize as follows,\footnote{~Since~$\tau_{Dk \ell}$ has non-trivial~$Sp(2N - 2, \Z)$ monodromy around the multi-monopole point, we must pick a branch of the logarithm to render the matrix~$t_{k\ell}$ in~\eqref{eq:tauddef} well defined. As explained below, we will mostly work with configurations~$a_{Dk}$ that are positive imaginary, so that~$-i a_{Dk} > 0$. We can then choose the principal branch of the logarithm, so that~$\log(- i a_{Dk})$ is real.}
\begin{equation}\label{eq:tauddef}
\tau_{D k \ell} =  {i} \left( -  \frac{1}{2\pi} \delta_{k\ell} \log \Big({-i a_{D k} \over \Lambda}\Big) + {2\pi} t_{k\ell}\right) + \CO(a_D)~, \qquad t_{k\ell} = t_{\ell k} \in \R~.
\end{equation}
Here~$t_{k\ell}$ is a real, symmetric matrix that accounts for the leading threshold corrections due to massive particles that have been integrated out in the low-energy effective description on the Coulomb branch. As such we will often refer to~$t_{k\ell}$ as the threshold matrix. Clearly the imaginary part of~\eqref{eq:tauddef}, which describes the matrix of gauge coupling constants at low energies, is positive definite as long as the~$a_{Dk}$ are sufficiently close to the multi-monopole point. Note that the off-diagonal elements of the matrix~$t_{k\ell}$ can be accessed by taking~$a_{Dk} \rightarrow 0$ in~\eqref{eq:tauddef}, since the corresponding~$\tau_{Dk\ell}$ has a finite limit.\footnote{~The physical importance of these off-diagonal terms was first stressed in~\cite{Douglas:1995nw}. They also play an important role in~\cite{FutureUs}.} By contrast, the diagonal matrix elements~$t_{kk}$ are finite threshold corrections to the divergent logarithms in~$\tau_{D kk}$. Thus computing them is more challenging; any such computation must regularize the logarithms by perturbing away from the multi-monopole point.  

As was emphasized in~\cite{Douglas:1995nw}, the structure of the threshold matrix~$t_{k\ell}$ encodes important information about the~$SU(N)$ $\CN=2$ gauge theory near the multi-monopole point -- in particular its massive spectrum there. Upon softly breaking~$\CN=2 \rightarrow \CN=1$ (as reviewed below~\eqref{eq:tdmatirlog}) the threshold matrix is needed to determine the spectrum of light particles, as well as the confining string tensions. Roughly speaking, this is due to the fact that~$t_{k\ell}$ is the matrix of gauge-kinetic terms in the low-energy~$U(1)_D^{N-1}$ gauge theory that couples to the~$N-1$ massless monopole hypermultiplets at the multi-monopole point.

Our primary interest in the threshold matrix~$t_{k\ell}$ comes from the recent observation~\cite{Cordova:2018acb} that the dynamics of non-supersymmetric adjoint QCD with gauge group $G$ and two adjoint quarks can be analyzed by adding a certain soft supersymmetry-breaking mass term for the adjoint scalars to the pure~$\CN=2$ supersymmetric gauge theory with the same gauge group $G$.\footnote{~In this embedding, the adjoint quarks are simply the two gauginos of the~$\CN=2$ gauge theory.} The case~$G = SU(2)$ was analyzed in~\cite{Cordova:2018acb}, where it was found that the expected confining and chiral-symmetry breaking phase of adjoint QCD emerged from the dynamics of the monopole and dyon points in the presence of the soft supersymmetry-breaking scalar mass. In upcoming work~\cite{FutureUs} we extend this to~$G = SU(N)$ for all~$N$, where the soft supersymmetry-breaking mass deformation leads to a rich structure of phases and phase transitions that can be analyzed by focusing on the multi-dyon points. This analysis crucially depends on the detailed properties of the threshold matrix~$t_{k\ell}$ in~\eqref{eq:tauddef}. 
 
A procedure for computing~$t_{k\ell}$ was outlined in~\cite{Douglas:1995nw}, where the authors considered a particular one-parameter family~$a_{Dk}(s)$ that approaches the multi-monopole point as~$s\rightarrow 0$. However, this procedure was ultimately only carried out for the elements of~$t_{k\ell}$ that dominate in the 't Hooft large-$N$ limit of the theory emphasized in~\cite{Douglas:1995nw}. Exact results for~$N=2$ and~$N = 3$ were obtained in~\cite{Klemm:1995wp}. Subsequently, the authors of~\cite{DHoker:1997mlo} developed a systematic method to compute higher-order corrections to~$\tau_{D k \ell}$ for all~$N$, starting with the~$\CO(a_D)$ terms in~\eqref{eq:tauddef}, but they did not compute~$t_{k\ell}$. A formula for the off-diagonal elements of~$t_{k\ell}$ was conjectured in~\cite{Edelstein:1999fz, Edelstein:1999tb}, and subsequently confirmed in~\cite{Edelstein:2000aj} (see also~\cite{Braden:2000he}), using the relationship of Seiberg-Witten theory to integrable hierarchies. More recently, the authors of~\cite{Bonelli:2017ptp} presented a computation of~$t_{k\ell}$ based on (partially conjectural) topological string and matrix model machinery. While their formula agrees with previous results for the off-diagonal part of~$t_{k\ell}$, they noted disagreements with previous statements about the diagonal part. See section~\ref{ssec:litcomp} below for further comments on the literature. 

In the supersymmetry-breaking analysis~\cite{FutureUs} we rely on the quantitative details of the threshold matrix~$t_{k\ell}$ -- not just its qualitative or large-$N$ features. For this reason we present a detailed and direct calculation of~$t_{k\ell}$ using standard Seiberg-Witten technology. As explained below~\eqref{eq:tauddef}, a full calculation of~$t_{k\ell}$ requires regularizing the logarithmic singularities in~\eqref{eq:tauddef} by perturbing away from the multi-monopole point. Here we will follow and extend the regularization method of~\cite{DHoker:1997mlo}, which we review in section~\ref{sec:setrev}.\footnote{~See section~\ref{ssec:litcomp} and appendix~\ref{app:DScomp} for more details on the regularization method used in~\cite{Douglas:1995nw}.}  

Our main result (derived in section~\ref{sec:acomp}) is a computation of the~$a_k$ periods near the multi-monopole point,\footnote{~Note that substituting~\eqref{eq:akanswerintro} into~\eqref{eq:tdaadmat} leads to~\eqref{eq:tauddef}.} 
\begin{equation}\label{eq:akanswerintro}
a_k\left(a_{D\ell}\right) = a_k\left(a_{D\ell} = 0\right) +  {i \over 2 \pi} a_{D k} \left(- \log {- i a_{Dk} \over \Lambda} + 1\right) +   {2\pi} i \sum_{\ell =1}^{N-1}  t_{k \ell} a_{D\ell}+ \CO(a^2_D)~.
\end{equation}
We find that the~$a_k$ periods at the multi-monopole point are given by\footnote{~These were first computed in~\cite{Douglas:1995nw} (see the discussion around~\eqref{eq:akatmmpi}). Here we include a scheme-dependent prefactor that depends on our normalization conventions for the strong-coupling scale~$\Lambda$. Our conventions are spelled out in section~\ref{ssec:swsolrev}, {and the differences between our conventions and those used in~\cite{Douglas:1995nw} are described in appendix~\ref{app:DScomp}.}}
\begin{equation}\label{eq:akadzansint}
a_k\left(a_{D\ell} = 0\right) = -{2 N\Lambda \over \pi} \sin {\pi k \over N}~, \qquad 
\end{equation}
while our result for the elements of the threshold matrix~$t_{k\ell}$ is given by
\begin{equation}\label{eq:tmatansint}
t_{kk} =\frac{1}{{4\pi^2}} \log \left(16 N \sin^3 {\pi k \over N}\right)~, \qquad t_{k\ell} = \frac{1}{{4\pi^2}} \log {{ \sin^2{(k+\ell) \pi \over 2 N } } \over { \sin^2{(k-\ell) \pi \over 2 N } } }\qquad (k \neq \ell)~.
\end{equation}
Note that in addition to the symmetry~$t_{k\ell} = t_{\ell k}$ that is necessarily present~(see~\eqref{eq:tauddef}), the threshold matrix also satisfies
\begin{equation}
t_{k \ell} = t_{N-k, N-\ell}~.
\end{equation}
This follows from the charge conjugation symmetry of the underlying~$SU(N)$ gauge theory, which is preserved at the multi-monopole point. (This will play an important role in~\cite{FutureUs}.) It is tempting to speculate that the special form of~$t_{k\ell}$ in~\eqref{eq:tmatansint} -- a logarithm of sine functions -- can be explained by appealing to the spectrum of heavy BPS states at the multi-monopole point, whose masses are determined by the $a_k \sim \sin {\pi k \over N}$ at that point (see~\eqref{eq:akadzansint}).\footnote{~An interpretation of~$t_{k\ell}$ in terms of the massive BPS spectrum near the multi-monopole point must contend with the fact that this point lies on a wall of marginal stability across which the BPS spectrum jumps \cite{Seiberg:1994rs, Douglas:1995nw}, while~$t_{k\ell}$ is wall-crossing invariant. This suggests an approach along the lines of~\cite{Gaiotto:2008cd}, where a similar puzzle was encountered and resolved.} However, we do not know of such an explanation.\footnote{~\label{ft:nicefootnote}As another possible hint, we record the following interesting, exact formula (inspired by~\cite{Douglas:1995nw} and appendix~\ref{app:DScomp}) for the off-diagonal elements of~$t_{k\ell}$ in~\eqref{eq:tmatansint}, 
\begin{equation}\label{eq:niceformula}
t_{k\ell} =\frac{1}{ {4\pi^2}} \log {{ \sin^2{(k+\ell) \pi \over 2 N } } \over { \sin^2{(k-\ell) \pi \over 2 N } } } = \frac{1}{ {4\pi^2}}  \sum_{p = 1}^\infty {4 \over p} \sin{p k \pi \over N} \sin{p\ell \pi \over N}~ \qquad (k \neq \ell)~.
\end{equation} 
To show this, we write the sum over~$p$ as~$\sum_{p = 1}^\infty {2 \over p} \left(\cos{p(k-\ell) \pi \over N} - \cos{p(k+\ell) \pi \over N}\right)$, which can be evaluated using~$\sum_{p = 1}^\infty {\cos{px} \over p} = - \log \left(2 \sin {x \over 2}\right)$, valid for~$x \in \R - 2 \pi \Z$. In turn, the latter formula follows from writing~$\cos px$ in terms of exponentials and using~$\sum_{p = 1}^\infty {z^p \over p} = - \log(1-z)$, with~$z = e^{\pm ix}$. 
}

\subsection{Comparison with the Literature}\label{ssec:litcomp}

In this subsection we compare our results~\eqref{eq:akanswerintro}, \eqref{eq:akadzansint}, and~\eqref{eq:tmatansint} to the existing literature in more detail. Along the way, we clarify some lingering inconsistencies. 

Using Picard-Fuchs equations, the authors of~\cite{Klemm:1995wp} found an expansion of the dual prepotential~$\CF_D(a_D)$ around the multi-monopole point for~$SU(2)$ and~$SU(3)$ gauge theories. The prepotential for~$SU(2)$ is given above equation (2.11) in~\cite{Klemm:1995wp}. From it we can compute~$a =  {\CF_D}'(a_D)$,\footnote{{~More precisely, the authors of~\cite{Klemm:1995wp} use~$a = - \CF_D'(a_D)$ and~$\tau_D = - \frac{da}{da_D}$, but the two minus signs cancel in~$\tau_D = \CF_D''(a_D)$. Thus our~$a$-periods differ from theirs by a sign, while the gauge couplings agree. The same comment applies to the~$SU(3)$ case described around~\eqref{eq:a1su3ex}.}}
\begin{equation}\label{eq:klemmsu2}
a(a_D) = - {2 \hat \Lambda \over \pi} + {i a_D \over 2 \pi} \left( - \log {- i a_D \over 16 \hat \Lambda} + 1\right) + \CO(a_D^2)~,
\end{equation} 
where we use~$\hat \Lambda$ to denote the strong coupling scale in the conventions of~\cite{Klemm:1995wp}. Comparing the constant term~$a(a_D = 0)$ in~\eqref{eq:klemmsu2} with~\eqref{eq:akadzansint}, we find agreement if~$\hat \Lambda = 2 \Lambda$. By comparing~\eqref{eq:klemmsu2} with~\eqref{eq:akanswerintro}, we then read off~${4\pi^2} t_{11} = \log 32$, in agreement with our result~\eqref{eq:tmatansint} for~$N=2$. 

In the~$SU(3)$ case the prepotential~$\CF_D(a_D)$ around the multi-monopole point is given in equations (6.13) and (6.14) of~\cite{Klemm:1995wp}. From it we can compute
\begin{equation}\label{eq:a1su3ex}
a_1 = {\d \CF_D\over \d a_{D1}} = - {2^{2/3}  \, 3 \sqrt 3 \, \hat \Lambda \over \pi} + {i \over 2 \pi} a_{D1} \left( - \log {- i a_{D1} \over 2^{5/3} \, 3^{5/2} \,  \hat \Lambda} + 1 \right) + {i \over 2 \pi} a_{D2} \log 4 + \CO(a_D^2)~, 
\end{equation}
and an analogous formula for~$a_2$, which can be obtained by exchanging~$a_{D1} \leftrightarrow a_{D2}$ in~\eqref{eq:a1su3ex}.\footnote{~Note that the prepotential in equations (6.13) and (6.14) of~\cite{Klemm:1995wp} is invariant under the charge-conjugation symmetry~$a_{D1} \leftrightarrow a_{D2}$.} Again we use~$\hat \Lambda$ to denote the strong coupling scale in the conventions of~\cite{Klemm:1995wp}. We proceed as above: by comparing the constant term~$a(a_D = 0)$ in~\eqref{eq:a1su3ex} with~\eqref{eq:akadzansint}, we find agreement if~$\hat \Lambda = 2^{-2/3} \Lambda$. Substituting back into~\eqref{eq:a1su3ex} and comparing with~\eqref{eq:akanswerintro}, we can then read off~${4\pi^2} t_{11} = \log 2 + {5 \over 2} \log 3$ and~${4\pi^2} t_{12} = \log 4$, in agreement with our result~\eqref{eq:tmatansint} for~$N=3$. 

We now compare our results to those of~\cite{Douglas:1995nw}, which apply to~$SU(N)$ gauge theories in the large-$N$ limit. In order to keep the present discussion brief, we defer a more detailed review of~\cite{Douglas:1995nw} to appendix~\ref{app:DScomp}, which also contains some new results (see below). As was already mentioned above, the authors of~\cite{Douglas:1995nw} considered a one-parameter scaling trajectory~$a_{Dk}(s)$ (with real parameter~$s$) that approaches the multi-monopole point as~$s\rightarrow 0$ (see appendix~\ref{app:DScomp}),\footnote{{~Here we  describe the results of~\cite{Douglas:1995nw} in our conventions; see appendix~\ref{app:DScomp} for further details.}}
\begin{equation}\label{eq:addsint}
a_{Dk}(s) = {2 i  \Lambda s \over N} \sin {\pi k \over N} + \CO(s^2)~.
\end{equation} 
Substituting this into~\eqref{eq:tauddef} and using our answer for the threshold matrix~$t_{k\ell}$ in~\eqref{eq:tmatansint}, we find that
\begin{equation}\label{eq:dsslice}
\tau_{D k \ell}(s) = {i \over 2 \pi} \begin{cases}
- \log s + \log \left(8 N^2  \sin^2 {\pi k \over N} \right) & \text{if} \quad k = \ell~, \\
\log {{ \sin^2{(k+\ell) \pi \over 2 N } } \over { \sin^2{(k-\ell) \pi \over 2 N } } } & \text{if} \quad k \neq \ell
\end{cases}~.
\end{equation}
We will now compare this answer to the calculations in~\cite{Douglas:1995nw}. Although the approach outlined there in principle allows one to calculate all $s$-independent terms in~\eqref{eq:dsslice}, the authors of~\cite{Douglas:1995nw} only explicitly evaluated those terms that grow without bound in the~$N \rightarrow \infty$ limit. As reviewed in appendix~\ref{app:DScomp}, it follows from the results of~\cite{Douglas:1995nw} that the elements of~$\tau_{D k \ell}(s)$ that have such growing large-$N$ contributions are\footnote{~As explained in appendix~\ref{app:DScomp}, \eqref{eq:dstaud} also applies when~$k = \ell$ if we omit the factor~$(k-\ell)^2$ in the logarithm.} 
\begin{equation}\label{eq:dstaud}
\tau_{D k\ell}(s) = {i \over 2 \pi} \left(- \delta_{k\ell} \log s + \log {N^2 \over (k-\ell)^2 } \right) + \CO(1) + \CO(s)~,  \quad k, \ell = \CO(N)~, \quad {k - \ell \over N} \rightarrow 0~.
\end{equation}
Here the~$\CO(1)$ terms in~$\tau_{D k\ell}(s)$ are constant as~$s \rightarrow 0$ and remain bounded at large~$N$. This precisely agrees with~\eqref{eq:dsslice} for those~$k, \ell$ indicated in~\eqref{eq:dstaud}.\footnote{~\label{ft:wrongds}Some formulas in~\cite{Douglas:1995nw} have subsequently been extrapolated beyond the regime in~\eqref{eq:dstaud}, where they no longer apply. For instance, the authors of~\cite{DHoker:1997mlo,Edelstein:1999fz,Edelstein:1999tb} appealed to~\cite{Douglas:1995nw} to argue that the diagonal elements~$t_{kk}$ of the threshold matrix are proportional to~$\log \sin{ \pi k \over N}$, rather than our result in~\eqref{eq:tmatansint}. Note that these two expressions do not agree in the large-$N$ limit.} It was argued in~\cite{Douglas:1995nw} that the~$\sim \log N^2$ threshold corrections in~\eqref{eq:dstaud} are due to light particles of mass~$\sim {\Lambda \over N^2}$, which impose a cutoff on the low-energy effective theory that vanishes in the large-$N$ limit. In appendix~\ref{app:DScomp} we show how to explicitly extend the computations in~\cite{Douglas:1995nw} to finite~$N$, and we recover the full answer in~\eqref{eq:dsslice}.

By combining elements of~\cite{Douglas:1995nw} with insights from integrable hierarchies, the authors of~\cite{Edelstein:1999fz, Edelstein:1999tb} conjectured an exact (but complicated) formula for the off-diagonal elements of the threshold matrix~$t_{k\ell}$.\footnote{~As was pointed out in footnote~\ref{ft:wrongds}, the diagonal elements~$t_{kk}$ are not correct in these papers.} A simpler expression for these off-diagonal elements was subsequently obtained in~\cite{Edelstein:2000aj}, where they were recomputed (again within the framework of integrable hierarchies) and used to numerically verify the conjecture of~\cite{Edelstein:1999fz, Edelstein:1999tb} for low values of~$N$. The off-diagonal elements in equations (6.11) and (6.12) of~\cite{Edelstein:2000aj} are easily seen to match our off-diagonal elements of~$\tau_{Dk\ell}$ in~\eqref{eq:tauddef} and~\eqref{eq:tmatansint}, as well as~\eqref{eq:dsslice}. The off-diagonal elements of~$\tau_{Dk\ell}$ were also examined in~\cite{Braden:2000he}, where they were expressed in a form (see their equation (169)) that exactly agrees with our~\eqref{eq:dsslice}, and shown to agree with the conjecture of~\cite{Edelstein:1999fz, Edelstein:1999tb}. 

The only complete result for the threshold matrix~$t_{k\ell}$ (including its diagonal elements) that we are aware of was recently put forward in~\cite{Bonelli:2017ptp}, using a dual matrix model that was motivated by appealing to conjectures in topological string theory. While the authors found agreement with~\cite{Edelstein:1999fz, Edelstein:1999tb, Edelstein:2000aj} for the off-diagonal elements of~$t_{k\ell}$, they also noted disagreement for the diagonal elements~$t_{kk}$. We will now compare the matrix-model results of~\cite{Bonelli:2017ptp} to ours. Their results are expressed in terms of a matrix-model (MM) prepotential~$\CF_D^\text{MM}(T_k)$, where the~$T_k$ are the dimensionless 't Hooft couplings of the matrix model, which are to be identified with the~$a_{Dk}$ periods (see equation (4.7) of~\cite{Bonelli:2017ptp}). We would like to convert to a prepotential~$\CF_D(a_D)$ from which we can compute~$a_k = {\d \CF_D / \d a_{Dk}}$ and compare to our formulas~\eqref{eq:akanswerintro}, \eqref{eq:akadzansint}, and \eqref{eq:tmatansint}. By examining the logarithmic terms, we are led to identify\footnote{~Note that our relation between~$T_k$ and~$a_{Dk}$ involves a factor of~$-i$ that is absent in equation (4.7) of~\cite{Bonelli:2017ptp}.}\begin{equation}\label{eq:mmvars}
T_k = {- i a_{Dk} \over \hat \Lambda}~, \qquad \CF_D(a_D) = {i \hat \Lambda^2 \over 2 \pi} \CF_D^\text{MM} (T_k)~.
\end{equation}
Here~$\hat \Lambda$ is a strong-coupling scale introduced for dimensional reasons, whose relation to our~$\Lambda$ will be fixed below.  By substituting the matrix-model prepotential in equations (4.14) and (4.15) of~\cite{Bonelli:2017ptp} into~\eqref{eq:mmvars}, we find that the results of~\cite{Bonelli:2017ptp} imply that
\begin{equation}\label{eq:akfrommm}
a_k = {\d \CF_D \over \d a_{Dk}} = - { \hat \Lambda \over 2\pi} \, { \sin{\pi k \over N} \over \sin {\pi \over N}} + {i a_{Dk} \over 2 \pi} \left( - \log {- i a_{Dk} \, \sin{\pi \over N} \over  4 \hat \Lambda \sin^3 {\pi k \over N} } + 1\right) + {i \over 2 \pi} \sum_{\ell \neq k} a_{D \ell} \log {\sin^2 {\pi (k + \ell) \over 2 N} \over \sin^2 {\pi (k - \ell) \over 2 N}}~.
\end{equation}
Comparing with~\eqref{eq:akanswerintro} and~\eqref{eq:tmatansint}, we see that the last term in~\eqref{eq:akfrommm} correctly accounts for the off-diagonal elements of our~$t_{k\ell}$ matrix. In order to find the scheme change that relates~$\hat \Lambda$ to our~$\Lambda$, we compare the constant term~$a(a_D = 0)$ in~\eqref{eq:akfrommm} with~\eqref{eq:akadzansint}, finding agreement if~$\hat \Lambda = 4 N \Lambda \sin{\pi \over N}$.\footnote{~Note that this is an~$N$-dependent change of scheme, though both~$\Lambda$ and~$\hat \Lambda$ are~$\CO(1)$ in the large-$N$ limit.} Substituting back into~\eqref{eq:akfrommm}, we see that the remaining terms correctly account for their counterparts in~\eqref{eq:akanswerintro} and~\eqref{eq:tmatansint}, including an exact match for the diagonal elements~$t_{kk}$ of our threshold matrix.

\section{Setup and Review}\label{sec:setrev}

\subsection{The $SU(N)$ Seiberg-Witten Solution}\label{ssec:swsolrev}

In this section we briefly review aspects of the Seiberg-Witten solution of the pure $SU(N)$ gauge theory, as determined in~\cite{Seiberg:1994rs,Klemm:1994qs,Argyres:1994xh}. The Seiberg-Witten curve $\Sigma$ is a hyperelliptic Riemann surface of genus $N-1$. It can be presented in many ways that are useful for various purposes. These presentations may differ by coordinate changes, as well as by an overall rescaling of the strong-coupling scale $\Lambda$ (i.e.~by a scheme change). Using this freedom, we can express the Seiberg-Witten curve in the following form,\begin{equation}\label{eq:swcurve}
y^2 = \big(C_N(x)\big)^2 - 1~.
\end{equation}
Here $x, y$ are dimensionless complex coordinates, while~$C_N(x)$ is a degree $N$ polynomial in $x$ whose dimensionless coefficients depend on the gauge-invariant Coulomb-branch order parameters~$u_n = \tr \phi^n = \sum_{i = 1}^N \phi_i^n$ (where~$\phi_i$ are the eigenvalues of~$\phi$),
\begin{equation}\label{eq:cdef}
C_N(x) = 2^{N-1} \det\left(x - {\phi \over 2 \Lambda}\right) = 2^{N-1} \prod_{i = 1}^N \left(x - {\phi_i \over 2 \Lambda} \right)~.
\end{equation}
Since $\tr \phi = \sum_{i = 1}^N \phi_i = 0$, the $\CO(x^{N-1})$ term in $C_N(x)$ vanishes, so that
\begin{equation}\label{eq:noxsq}
C_N(x) = 2^{N-1} \left(x^N - {u_2 \over 8 \Lambda^2} \, x^{N-2} + \cdots\right)~,
\end{equation} 
where the ellipsis denotes terms of order $x^{N-3}$ or lower in~$x$.

The Seiberg-Witten differential (which has mass-dimension one) is given by
\begin{equation}\label{eq:swdiff}
\lambda = (2 \Lambda) \, {x C'_N(x) d x \over y} =(2 \Lambda) \, {x C'_N(x) d x \over \sqrt{\big(C_N(x)\big)^2 - 1}}~.
\end{equation}
It is a meromorphic one-form on $\Sigma$. Once we fix a set of a canonical $A$- and $B$-cycles on $\Sigma$, we can determine the special Coulomb-branch coordinates $a_k$ and $a_{Dk}$ by integrating $\lambda$ over these cycles,
\begin{equation}\label{eq:aaddef}
a_k = {1 \over 2 \pi i} \oint_{A_k} \lambda~, \qquad a_{D k} = {1 \over 2 \pi i} \oint_{B_k} \lambda \qquad (k = 1, \ldots, N-1)~.
\end{equation} 
Different choices of $A$- and $B$-cycles lead to special coordinates that differ by $Sp(2N - 2, \Z)$ electric-magnetic duality transformations of the low-energy $U(1)^{N-1}$ gauge theory on the Coulomb branch. 

Unless stated otherwise, we set the strong-coupling scale $\Lambda$ (which is the only dimensionful parameter in the problem) to $\Lambda = \half$, so that the Seiberg-Witten differential~\eqref{eq:swdiff} simplifies to
\begin{equation}\label{eq:swdiffnoL}
\lambda = {x \, C'_N(x) \, dx \over \sqrt{\big(C_N(x)\big)^2 -1}}~.
\end{equation}

\subsection{The Multi-Monopole Point}

\begin{figure}[htb]
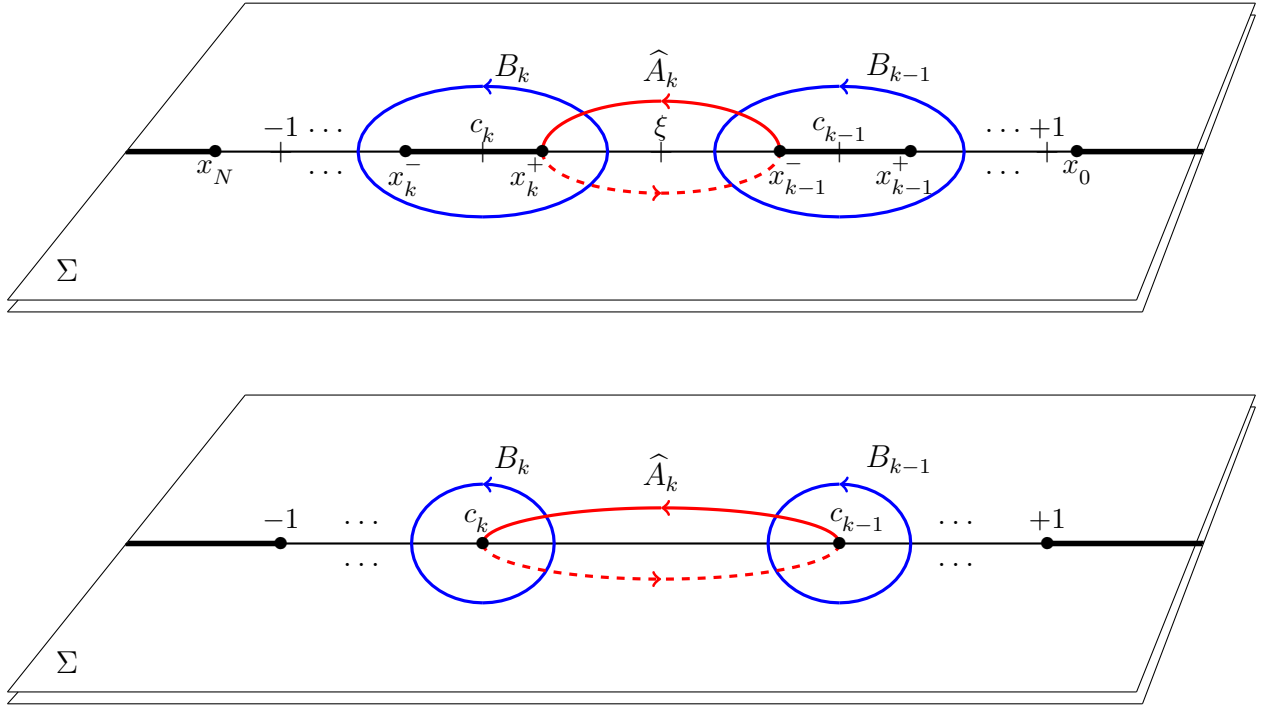

\begin{center}
\tikzpicture[scale=0.79]
\scope[xshift=0cm,yshift=0cm]
\draw [thick] (-3,0) -- (15,0);
\draw [very thick] (-2.97,0.02) -- (-0.4,0.02);
\draw [very thick] (-3.015,-0.02) -- (-0.4,-0.02);
\draw [very thick] (12.5,0.02) -- (15.13,0.02);
\draw [very thick] (12.5,-0.02) -- (15.11,-0.02);
\draw  [blue, very thick, domain=90:270] plot ({3+1.2*cos(\x)},{sin(\x)});
\draw  [blue, very thick, domain=270:450,->] plot ({3+1.2*cos(\x)},{sin(\x)});
\draw  [blue, very thick, domain=90:270] plot ({9+1.2*cos(\x)},{sin(\x)});
\draw  [blue, very thick, domain=270:450,->] plot ({9+1.2*cos(\x)},{sin(\x)});
\draw (-0.4,0) node{$\bullet$};
\draw (12.5,0.4) node{\small $+1$};
\draw (-0.4,0.4) node{\small $-1$};
\draw (3.5,1.4) node{$B_k$};
\draw (10,1.4) node{$B_{k-1}$};
\draw (6,1.2) node{$\hat A_k$};
\draw  [red, very thick, domain=0:90,->] plot ({6+3*cos(\x)},{0.6*sin(\x)});
\draw  [red, very thick, domain=90:180] plot ({6+3*cos(\x)},{0.6*sin(\x)});
\draw  [red, very thick, dashed, domain=180:270,->] plot ({6+3*cos(\x)},{0.6*sin(\x)});
\draw  [red, very thick, dashed, domain=360:270] plot ({6+3*cos(\x)},{0.6*sin(\x)});
\draw (3,0) node{$\bullet$};
\draw (9,0) node{$\bullet$};
\draw (12.5,0) node{$\bullet$};
\draw (1,0.35) node{$\dots$};
\draw (1,-0.35) node{$\dots$};
\draw (11,0.35) node{$\dots$};
\draw (11,-0.35) node{$\dots$};
\draw (2.9,0.38) node{\small $c_k$};
\draw (9.3,0.38) node{\small $c_{k-1}$};
\draw (-5,-2.5) -- (14,-2.5);
\draw (-1,2.5) -- (16,2.5);
\draw (-5,-2.5) -- (-1,2.5);
\draw (14,-2.5) -- (16,2.5);
\draw (-5,-2.7) -- (14.1,-2.7);
\draw (16,2.3) -- (14.1,-2.7);
\draw (16,2.3) -- (15.91, 2.3);
\draw (-5,-2.7) -- (-4.82, -2.5);
\draw (-4,-2) node{$\Sigma$};
\endscope
\scope[xshift=0cm,yshift=6.6cm]
\draw [thick] (-3,0) -- (15,0);
\draw [very thick] (-2.97,0.02) -- (-1.5,0.02);
\draw [very thick] (-3.015,-0.02) -- (-1.5,-0.02);
\draw [very thick] (13,0.02) -- (15.13,0.02);
\draw [very thick] (13,-0.02) -- (15.11,-0.02);
\draw [very thick] (1.7,0.02) -- (4,0.02);
\draw [very thick] (1.7,-0.02) -- (4,-0.02);
\draw [very thick] (8,0.02) -- (10.2,0.02);
\draw [very thick] (8,-0.02) -- (10.2,-0.02);
\draw  [blue, very thick, domain=90:270] plot ({3+2.1*cos(\x)},{1.1*sin(\x)});
\draw  [blue, very thick, domain=270:450,->] plot ({3+2.1*cos(\x)},{1.1*sin(\x)});
\draw  [blue, very thick, domain=90:270] plot ({9+2.1*cos(\x)},{1.1*sin(\x)});
\draw  [blue, very thick, domain=270:450,->] plot ({9+2.1*cos(\x)},{1.1*sin(\x)});
\draw (13,-0.35) node{\small $x_0$};
\draw (12.5,0.4) node{\small $+1$};
\draw (-0.4,0.4) node{\small $-1$};
\draw (1.7,-0.42) node{\small $x_k^-$};
\draw (3.75,-0.42) node{\small $x_k^+$};
\draw (3.5,1.4) node{$B_k$};
\draw (8.3,-0.42) node{\small $x_{k-1}^-$};
\draw (10.1,-0.42) node{\small $x_{k-1}^+$};
\draw (-1.5,-0.35) node{\small $x_N$};
\draw (10,1.4) node{$B_{k-1}$};
\draw (6,1.4) node{$\hat A_k$};
\draw  [red, very thick, dashed, domain=180:270,->] plot ({6+2*cos(\x)},{0.7*sin(\x)});
\draw  [red, very thick, dashed, domain=270:360] plot ({6+2*cos(\x)},{0.7*sin(\x)});
\draw  [red, very thick, domain=180:90] plot ({6+2*cos(\x)},{0.85*sin(\x)});
\draw  [red, very thick, domain=0:90,->] plot ({6+2*cos(\x)},{0.85*sin(\x)});
\draw (-1.5,0) node{$\bullet$};
\draw (1.7,0) node{$\bullet$};
\draw (4,0) node{$\bullet$};
\draw (8,0) node{$\bullet$};
\draw (10.2,0) node{$\bullet$};
\draw (13,0) node{$\bullet$};
\draw (3,0) node{$+$};
\draw (9,0) node{$+$};
\draw (6,0) node{$+$};
\draw (12.5,0) node{$+$};
\draw (-0.4,0) node{$+$};
\draw (0.4,0.35) node{$\dots$};
\draw (0.4,-0.35) node{$\dots$};
\draw (11.8,0.35) node{$\dots$};
\draw (11.8,-0.35) node{$\dots$};
\draw (3,0.35) node{\small $c_k$};
\draw (9,0.35) node{\small $c_{k-1}$};
\draw (6,0.4) node{\small $\xi$};
\draw (-5,-2.5) -- (14,-2.5);
\draw (-1,2.5) -- (16,2.5);
\draw (-5,-2.5) -- (-1,2.5);
\draw (14,-2.5) -- (16,2.5);
\draw (-5,-2.7) -- (14.1,-2.7);
\draw (16,2.3) -- (14.1,-2.7);
\draw (16,2.3) -- (15.91, 2.3);
\draw (-5,-2.7) -- (-4.82, -2.5);
\draw (-4,-2) node{$\Sigma$};
\endscope
\endtikzpicture
\bigskip
\caption{The figure in the lower panel represents the singular hyperelliptic Seiberg-Witten curve at the multi-monopole point, with branch cuts $(-\infty, -1) \cup (+1,+\infty)$, singular points $c_1, \ldots,  c_{N-1}$ (where the two Riemann sheets touch), and a choice of homology basis. The figure in the upper panel represents the regular hyperelliptic Seiberg-Witten curve away from the multi-monopole points, with branch cuts $(-\infty, x_N) \cup (x_{N-1}^-, x_{N-1}^+) \cup \cdots \cup (x_1^-, x_1^+) \cup (x_0,\infty)$, and a choice of homology basis that degenerates to the homology basis of the singular curve in the lower figure. Each branch cut $(x_k^-, x_k^+)$ of the regular curve degenerates to the corresponding singular point $c_k$ of the multi-monopole curve. For later use in subsection~\ref{ssec:atkgeq2}, an arbitrary  point $\xi \in (x_{k}^+, x_{k-1}^-)$ that is well separated from the endpoints of the interval has also been indicated.} \label{fig:1}
\end{center}
\end{figure}

As we reviewed in section~\ref{ssec:mmp}, there are $N$ multi-dyon points on the Coulomb branch of the $SU(N)$ gauge theory, and we focus on the multi-monopole point. It was shown in~\cite{Douglas:1995nw} that this point occurs when $C_N(x)$ in~\eqref{eq:swcurve} and \eqref{eq:cdef} is given by a degree $N$ Chebyshev polynomial,\footnote{~This definition of the Chebyshev polynomials is valid for $-1 \leq x \leq 1$, but it can be analytically continued to all $x \in \C$.}
\begin{equation}\label{eq:chebdef}
C_N(x)\big|_\text{multi-monopole} \equiv C_N^{(0)}(x) = \cos(N \arccos x)~. 
\end{equation}
Here and throughout the paper we use the superscript $(0)$ to denote quantities evaluated at the multi-monopole point. The leading terms in $C_N^{(0)}(x)$ are given by
\begin{equation}\label{eq:cheblead}
C_N^{(0)}(x) = 2^{N-1} \left(x^N - {N \over 4} x^{N-2} + \cdots\right)~,
\end{equation}
in accord with the general form of $C_N(x)$ in~\eqref{eq:noxsq}. By differentiating \eqref{eq:chebdef} we can derive a useful functional relation obeyed by $C_N^{(0)}(x)$ and its first derivative $C_N^{(0)}(x)'$,
\begin{equation}\label{eq:ccprel}
\big(C_N^{(0)}(x)\big)^2 - 1 = {x^2 - 1 \over N^2} \big(C_N^{(0)}(x)'\big)^2~.
\end{equation}
This relation can be used to analyze the branch and singular points of the Seiberg-Witten curve~\eqref{eq:swcurve} at the multi-monopole point, which occur when $y^2 = \left(C_N^{(0)}(x)\right)^2 - 1$ vanishes. To this end we use the following product representation for $C_N^{(0)}(x)'$,\footnote{~This formula can be derived by using the defining relation~\eqref{eq:chebdef} for $C_N^{(0)}(x)$ to argue that the $N-1$ zeroes of $C_N^{(0)}(x)' $ must be at $x = c_k = \cos\left(k \pi \over N\right)$, and then fixing the overall coefficient by comparing with~\eqref{eq:cheblead}.}
\begin{equation}\label{eq:cpprod}
C_N^{(0)}(x)' =  2^{N-1}N \prod_{k = 1}^{N-1} (x - c_k)~.
\end{equation}
Here, and for future use, we define the following shorthands,
\begin{equation}\label{eq:ckskdef}
c_k = \cos\left({k \pi \over N}\right)~, \qquad s_k = \sin \left({k \pi \over N}\right)~, \qquad k \in \Z~.
\end{equation}
Note that~$k$ can be any integer, though it will typically lie in the range $1 \leq k \leq N-1$.  Substituting~\eqref{eq:cpprod} into~\eqref{eq:ccprel}, we see that $y^2$ has $N-1$ double zeroes at $x = c_k$ and two simple zeroes at $x = \pm 1$. The simple zeroes correspond to non-singular branch points of the curve, while the double zeroes indicate that the curve has $N-1$ singular degeneration points reflecting the $N-1$ massless monopoles, as represented in the lower panel of figure~\ref{fig:1}. 

\medskip

As we will review in section~\ref{ssec:nearmmp} below, the branch cuts of the non-singular Seiberg-Witten curve in the vicinity of the multi-monopole point can be chosen so that the singular points~$x = c_k$ of the multi-monopole curve~$y^2 = \left(C_N^{(0)}(x)\right)^2 - 1$ arise from $N-1$ branch cuts that collapse to zero length. The only remaining branch cuts of the multi-monopole curve run from~$+1$ to~$+\infty$ and from~$-1$ to~$-\infty$ along the real axis (see figure~\ref{fig:1}). Up to an overall choice of sign (which amounts to a choice of Riemann sheet on the Seiberg-Witten curve), this specification of the branch cuts allows us to define the square root of~\eqref{eq:ccprel} as a well-defined, holomorphic function on the cut~$x$-plane (i.e.~$x \in \C - \{(-\infty, -1) \cup (1, \infty)\}$). We choose the overall sign so that the following identity holds,
\begin{equation}\label{eq:singsqrt}
\sqrt{\big(C_N^{(0)}(x)\big)^2 - 1} = - {i \over N} \sqrt{1-x^2} \, C_N^{(0)}(x)'~, \qquad -1 \leq x \leq 1~.
\end{equation}
In other words, the sign of the square root on the left-hand side varies with the sign of the polynomial~$C_N^{(0)}(x)'$. The identity~\eqref{eq:singsqrt} extends to the entire cut~$x$-plane, on which both sides are holomorphic functions defined by analytic continuation. Throughout the remainder of the paper we will define all square roots we encounter by ensuring compatibility with~\eqref{eq:singsqrt}.  

\subsection{The Vicinity of the Multi-Monopole Point}\label{ssec:nearmmp}

In order to explore the neighborhood of the multi-monopole point, we deform~\eqref{eq:chebdef} by adding to the Chebyshev polynomial $C_N^{(0)}(x)$ a degree $(N-2)$ polynomial $P_{N-2}(x)$,\footnote{~Recall from~\eqref{eq:noxsq} that the $\CO(x^N)$ and $\CO(x^{N-1})$ terms in~$C_N(x)$ are fixed (and in particular, that the latter vanishes).}
\begin{equation}\label{eq:cc0p}
C_N(x) = C_N^{(0)}(x) + P_{N-2}(x)~.
\end{equation}
The~$N-1$ complex coefficients of $P_{N-2}(x)$ describe the $N-1$ Coulomb branch directions along which we can approach the multi-monopole point by taking these coefficients to be sufficiently small (we will make this precise below). It is convenient to trade these $N-1$ coefficients for the values $P_k$ of $P_{N-2}(x)$ at $N-1$ distinct points, which we take to be $x = c_k$,
\begin{equation}
P_k = P_{N-2}(c_k)~, \qquad (k = 1, \ldots, N-1)~.
\end{equation} 
Conversely, we can express $P_{N-2}(x)$ in terms of the constants $P_k$ using the Lagrange interpolation formula, which we can in turn write in terms of the Chebyshev polynomial $C_N^{(0)}(x)$,
\begin{equation}\label{eq:lagrange}
P_{N-2}(x) = \sum_{k = 1}^{N-1} P_k \prod_{\substack{\ell = 1 \\ \ell \neq k}}^{N-1} {x - c_\ell \over c_k -c_\ell} = \sum_{k = 1}^{N-1} {P_k\, C_N^{(0)}(x)' \over (x-c_k) \, C_N^{(0)}(c_k)''}~.
\end{equation}

The addition of~$P_{N-2}(x)$ in~\eqref{eq:cc0p} deforms the zeroes of the curve~$y^2 = \left(C_N(x)\right)^2-1$. Recall from the discussion below~\eqref{eq:ckskdef} that the singular curve $y^2 = \left(C_N^{(0)}(x)\right)^2 -1$ at the multi-monopole point has simple zeroes at~$x = \pm 1$ and double zeroes at $x = c_k$. The effect of $P_{N-2}(x)$ is to shift the location of the simple zeroes, while the double zeroes split into pairs of simple zeroes. Explicitly, and to leading order in~$P_{N-2}(x)$, the zeroes of the Seiberg-Witten curve occur at the following values of $x$,\footnote{~\label{ft:zeroes}To see this, we approximate the curve as $y^2 \simeq \left(C_N^{(0)}(x)\right)^2 - 1 + 2 C_N^{(0)}(x) P_{N-2}(x)$ and use the identity~\eqref{eq:ccprel} to expand~$\left(C_N^{(0)}(x)\right)^2 - 1$ around its zeroes at~$x = c_k, \pm 1$. This requires evaluating~$C_N^{(0)}(x)$ and its first two derivatives at these zeroes, which can be done using the defining relation~\eqref{eq:chebdef}.}
\begin{equation}\label{eq:swzeroes}
\begin{split}
& x_0 = 1- \delta_0~, \hskip38pt  \delta_0 = {P_{N-2}(1) \over N^2}~, \\
& x_k^\pm = c_k \pm \delta_k~,   \hskip32pt \delta_k^2 = {(-1)^k 2  s_k^2 P_k \over N^2}~, \qquad (k = 1, \ldots, N-1)~,\\
& x_N = -1 + \delta_N~, \qquad  \delta_N = {(-1)^N P_{N-2}(-1) \over N^2}~.
\end{split}
\end{equation}
If all $P_k$ are non-zero, every one of these zeroes is simple and corresponds to a branch point of the (everywhere non-singular) Seiberg-Witten curve. We choose the branch cuts to run from~$+ \infty$ to $x_0$, from $x^+_k$ to~$x^-_k$, and from $x_N$ to $-\infty$ in the complex $x$-plane, as shown in figure~\ref{fig:1}.

If we scale towards the multi-monopole point by taking all $P_k \rightarrow 0$, then all $\delta$'s in~\eqref{eq:swzeroes} vanish, i.e.~the simple zeroes at $x_0, x_N$ approach~$+1, -1$ respectively, while the branch cuts connecting the simple zeroes $x^+_k$ and~$x^-_k$ collapse to singular double zeroes at $c_k$. The length of these cuts tracks the vanishing monopole masses as we approach the multi-monopole point (see section~\ref{ssec:adreview} below). Below, we will always choose the $P_k$ to be non-vanishing, but sufficiently small to ensure that the cuts from $x^+_k$ to $x^-_k$ (whose length is $2 |\delta_k|$) are much shorter than their distance to the nearest branch point.  

We will evaluate the special Coulomb-branch coordinates~$a_k$ and~$a_{Dk}$ as a function of the~$P_k$, to leading order in small $P_k$, by explicitly integrating the Seiberg-Witten differential~$\lambda$ in~\eqref{eq:swdiffnoL} over suitable~$A$- and $B$-cycles (specified below) as in~\eqref{eq:aaddef}. Since~$\lambda$ is a holomorphic one-form, the periods $a_k$ and $a_{Dk}$ are locally holomorphic functions of the~$P_k$. (Globally they may be branched and can undergo monodromy.) We can therefore simplify our computations by taking the~$P_k = (-1)^k |P_k|$ to be small real numbers of alternating sign, so that the~$\delta_k$ in~\eqref{eq:swzeroes} are small, real, and positive, i.e.~$\delta_k > 0$. Using~\eqref{eq:lagrange} we can further check that these sign choices imply~$\delta_0, \delta_N < 0$. In summary,
\begin{equation}\label{eq:signs}
P_k = (-1)^k |P_k|~, \qquad \delta_k > 0 \quad (k = 1, \ldots, N-1)~, \qquad \delta_0 < 0~, \qquad \delta_N < 0~.
\end{equation}
This leads to the simplified cut complex $x$-plane depicted in the upper panel of figure~\ref{fig:1}, since all branch cuts now run along the real axis.
 
Before we can compute the~$a_k$ and~$a_{Dk}$ periods we must choose a set of canonical~$A$- and $B$-cycles. Since we would like to associate the~$a_{Dk}$ with the light monopoles, we choose the cycles~$B_k~(k = 1, \ldots, N-1)$ to encircle the short branch cuts connecting~$x^\pm_k~$ once, in the counterclockwise direction (see figure~\ref{fig:1}).\footnote{~\label{ft:dscycle1}This matches the conventions of the $B_k$ cycles in~\cite{DHoker:1997mlo}, which agree with the $\alpha_k$ cycles in~\cite{Douglas:1995nw}.} Note that these cycles do not cross any branch cuts, so that the~$a_{D}$-periods~$a_{Dk} = {1 \over 2 \pi i} \oint_{B_k} \lambda$ in~\eqref{eq:aaddef} can be evaluated on a single sheet. This computation was carried out in~\cite{DHoker:1997mlo} and will be reviewed in section~\ref{ssec:adreview}. 

In order to define a suitable basis of $A$-cycles~$A_k~(k = 1, \ldots, N-1)$ conjugate to the~$B_k$ defined above, we first define a simpler basis~$\hat A_k$ of one-cycles that encircle the first~$N-1$ pairs of branch points in a counterclockwise direction, i.e.~$\hat A_1$ encircles~$x_0$ and~$x^+_1$, while~$\hat A_k~(k = 2, \ldots, N-1)$ encircles~$x_{k-1}^-$ and~$x_k^+$.\footnote{~\label{ft:dscycle2}These~$\hat A_k$ agree with the~$\gamma_k$ cycles in~\cite{Douglas:1995nw}.} (Note that the cycle~$\hat A_N$ encircling the final pair of branch points~$x_{N-1}^-$ and~$x_N$ is not linearly independent since~$\sum_{k=1}^N \hat A_k = 0$.) The way in which the~$\hat A_k$ cycles traverse the first and second sheets, as well as their intersections with the~$B_k$ cycles, are shown in figure~\ref{fig:1}. Explicitly, $\hat A_1$ intersects~$B_1$ in a negative sense, while~$\hat A_k~(k = 2, \ldots, N-1)$ intersects~$B_{k-1}$ in a positive sense and~$B_k$ in a negative sense. Thus the~$\hat A_k$ cycles are not themselves conjugate to the~$B_k$ cycles. However they can be used to construct a basis of conjugate~$A_k$ cycles as follows,
\begin{equation}\label{eq:acycles}
A_k  = \sum_{\ell = 1}^k \hat A_\ell~.
\end{equation}
This cycle intersects~$B_k$ in a negative sense, without intersecting any of the other~$B$-cycles.\footnote{~\label{ft:dscycle3}Our~$A_k$ cycles agree with the~$\beta_k$ cycles of~\cite{Douglas:1995nw}.} We can therefore compute the~$a$-periods~$a_k = {1 \over 2 \pi i} \oint_{A_k} \lambda$ in~\eqref{eq:aaddef} as follows,
\begin{equation}\label{eq:afromah}
a_k  = \sum_{\ell = 1}^k \hat a_\ell~, \qquad \hat a_k = {1 \over 2 \pi i} \oint_{\hat A_k} \lambda \qquad (k = 1, \ldots, N-1)~.
\end{equation}
The computation of the~$\hat a_k$, and hence the~$a_k$, will be described in section~\ref{sec:acomp}.

\subsection{Rewriting the Seiberg-Witten Differential}\label{ssec:swdiff}

The expansion of the~$a$- and~$a_D$-periods around the multi-monopole point is substantially complicated by the fact that the point around which we are expanding is singular. Following~\cite{DHoker:1997mlo}, this problem can be alleviated by a judicious rewriting of the Seiberg-Witten differential, which involves stripping off a locally exact one-form. To this end, we first introduce a family~$C_N^{(\mu)}(x)$ of degree $N$ polynomials that linearly interpolate between the Chebyshev polynomials~$C_N^{(0)}(x)$ and the polynomial~$C_N^{(1)} = C_N(x)$ in \eqref{eq:cc0p},
\begin{equation}\label{eq:cmudef}
C_N^{(\mu)}(x) = C_N^{(0)} + \mu P_{N-2}(x)~, \qquad 0 \leq \mu \leq 1~.
\end{equation}
We can then decompose the Seiberg-Witten differential~$\lambda$ in~\eqref{eq:swdiffnoL} as follows,
\begin{equation}\label{eq:lamdecomp}
\lambda = \t \lambda + dS~.
\end{equation}
Here~$\t \lambda$ is a locally defined meromorphic one-form given by the parametric integral 
\begin{equation}\label{eq:ltdef}
\t \lambda = -\int_0^1 d\mu \, {P_{N-2}(x) dx \over \sqrt{\big(C_N^{(\mu)}(x)\big)^2 - 1}}~,
\end{equation}
while~$S$ is a locally defined scalar function,
\begin{equation}\label{eq:sdef}
S(x) = x \log \left({C^{(1)}_N(x) + \sqrt{\big(C_N^{(1)}(x)\big)^2 - 1} \over C^{(0)}_N(x) + \sqrt{\big(C_N^{(0)}(x)\big)^2 - 1}}\right) - i N \sqrt{1-x^2}~.
\end{equation}
Neither~$\t \lambda$ nor~$S$ are globally well defined on the Seiberg-Witten curve. (In particular, $\t \lambda$ is not a valid Seiberg-Witten differential.) The reason is that both~$\t \lambda$ and~$S$ involve functions whose branch points do not coincide with the zeroes~\eqref{eq:swzeroes} of the Seiberg-Witten curve. We must therefore carefully define the branch cuts of the functions appearing in~\eqref{eq:ltdef} and~\eqref{eq:sdef}, which we will do below. 

Let us outline the derivation of the decomposition~\eqref{eq:lamdecomp}. It is straightforward to verify the identity
\begin{equation}\label{eq:dmuid}
{\d \over \d \mu} \left({x C^{(\mu)}_N(x)' dx \over \sqrt{\big(C^{(\mu)}_N(x)\big)^2-1}}\right) = - {P_{N-2}(x) dx \over\sqrt{\big(C^{(\mu)}_N(x)\big)^2-1}} +  d \left({x P_{N-2}(x) \over \sqrt{\big(C^{(\mu)}_N(x)\big)^2-1}}\right)~.
\end{equation}
Integrating from~$\mu = 0$ to~$\mu = 1$ and recalling the definitions of~$\lambda, \t \lambda$ in~\eqref{eq:swdiffnoL}, \eqref{eq:ltdef} gives
\begin{equation}\label{eq:intmuid}
\lambda - {x C^{(0)}_N(x)' dx \over \sqrt{\big(C_N^{(0)}(x)\big)^2-1}} = \t \lambda + d \left(\int_0^1 d\mu\, {x P_{N-2}(x) \over \sqrt{\big(C^{(\mu)}_N(x)\big)^2-1}}\right)~.
\end{equation}
Note that the second term on the left-hand side is (minus) the Seiberg-Witten differential of the multi-monopole curve. 

We pause here to discuss the branch cuts of the functions appearing in~\eqref{eq:dmuid} and~\eqref{eq:intmuid}. The zeroes of the function~$\big(C^{(\mu)}_N(x)\big)^2-1$ can be obtained from the zeroes of~$\big(C^{(1)}_N(x)\big)^2-1$ in~\eqref{eq:swzeroes} via a rescaling of~$P_{N-2}(x)$ by~$\mu$. Explicitly, they occur at the following values of~$x$,
\begin{equation}\label{eq:xmuzeroes}
x_0^{(\mu)} = 1 - \mu \delta_0~, \qquad x_k^{(\mu)\pm} = c_k \pm \sqrt{\mu} \delta_k \quad (k = 1, \ldots, N-1)~, \qquad x_N^{(\mu)} = -1 + \mu \delta_N~,
\end{equation}
with the~$\delta$'s given in~\eqref{eq:swzeroes}. We therefore choose the branch cuts of~$\sqrt{\big(C^{(\mu)}_N(x)\big)^2-1}$ in direct analogy with those of~$\sqrt{\big(C^{(1)}_N(x)\big)^2-1}$ (see the discussion below~\eqref{eq:swzeroes}), i.e.~running from~$+\infty$ to~$x_0^{(\mu)}$, from~$x_k^{(\mu)+}$ to ~$x_k^{(\mu)-}$, and from~$x_N^{(\mu)}$ to~$-\infty$ in the complex~$x$-plane. As~$\mu$ varies, these cuts continuously interpolate between those of the singular multi-monopole curve at~$\mu = 0$ and those of the non-singular Seiberg-Witten curve of interest at~$\mu = 1$ (see both panels of figure~\ref{fig:1}).

We now continue to simplify~\eqref{eq:intmuid}, starting with the Seiberg-Witten differential of the multi-monopole curve on the left-hand side. Using~\eqref{eq:singsqrt}, we obtain 
\begin{equation}\label{eq:singswdiff}
{x C^{(0)}_N(x)' dx \over \sqrt{\big(C_N^{(0)}(x)\big)^2-1}} = {i N x dx \over \sqrt{1-x^2}} = d\left(-iN \sqrt{1-x^2}\right)~, \qquad |x|< 1~. 
\end{equation}
Note that our choice of branch cuts in~\eqref{eq:singsqrt} implies that the branch cuts of~\eqref{eq:singswdiff} run from~$\pm1$ to~$\pm \infty$, with no branch cut between~$-1$ and~$1$. Finally, we can carry out the definite~$\mu$ integral in~\eqref{eq:intmuid} by changing variables to~$\t \mu = C_N^{(0)} + \mu P_{N-2}$, 
\begin{equation}\label{eq:defmuint}
\int_0^1 d\mu\, {x P_{N-2}(x) \over \sqrt{\big(C^{(\mu)}_N(x)\big)^2-1}} = x \int_{C_N^{(0)}}^{C_N^{(1)}} {d \t \mu \over \sqrt{ \t \mu^2 - 1}} = x \log \left({C^{(1)}_N(x) + \sqrt{\big(C_N^{(1)}(x)\big)^2 - 1} \over C^{(0)}_N(x) + \sqrt{\big(C_N^{(0)}(x)\big)^2 - 1}}\right)~.
\end{equation}
It follows from the discussion below~\eqref{eq:xmuzeroes} that the branch cuts of this function lie entirely within the intervals~$(1, +\infty)$, $(x_k^-, x_k^+)$, and~$(-\infty, -1)$. 

Substituting~\eqref{eq:singswdiff} and~\eqref{eq:defmuint} into~\eqref{eq:intmuid}, we obtain the decomposition~$\lambda = \t \lambda + dS$ in~\eqref{eq:lamdecomp}, with~$\t \lambda$ and~$S$ as defined in~\eqref{eq:ltdef} and~\eqref{eq:sdef}. Along the way we have seen that~$\t \lambda$ and~$S$ are not globally well defined on the Seiberg-Witten curve. They can however be defined in the cut $x$-plane, and we have chosen the cuts to lie entirely inside the intervals~$(1, +\infty)$, $(x_k^-, x_k^+)$ and~$(-\infty, -1)$. Most of our calculations below will stay away from these cuts. An exception occurs in section~\ref{ssec:a1scomp}.

\subsection{Expanding the $a_D$-Periods Around the Multi-Monopole Point}\label{ssec:adreview}

We now review the computation of the periods~$a_{Dk} = {1 \over 2 \pi i} \oint_{B_k} \lambda$ in~\eqref{eq:aaddef} to leading order in small~$P_{N-2}(x)$, as described in~\cite{DHoker:1997mlo}, where the calculation was also carried out to higher orders. Along the~$B_k$-cycles the one-form~$\t \lambda$ and the scalar function~$S$ in~\eqref{eq:ltdef} and~\eqref{eq:sdef} are single valued. (Recall the discussion at the end of section~\ref{ssec:swdiff}.) We are therefore free to use the decomposition~$\lambda = \t \lambda + dS$ in~\eqref{eq:lamdecomp} and drop the exact term~$dS$ in the computation of~$a_{Dk}$. Substituting the explicit form of~$\t \lambda$ in~\eqref{eq:ltdef}, and working to leading order in~$P_{N-2}(x)$, we thus find
\begin{equation}\label{eq:adsimp}
a_{Dk} = {1 \over 2 \pi i} \oint_{B_k} \t \lambda = - {1 \over 2 \pi i} \oint_{B_k} {P_{N-2}(x) dx \over \sqrt{\big(C_N^{(0)}(x)\big)^2-1}}~.
\end{equation} 
We now use~\eqref{eq:singsqrt} to simplify the square root in the denominator, 
\begin{equation}\label{eq:simpsqrt}
{1 \over \sqrt{\big(C_N^{(0)}(x)\big)^2-1}} = {i N \over C_N^{(0)}(x)' \sqrt{1-x^2}}~.
\end{equation}
Since~$C_N^{(0)}(x)'$ has simple zeroes at~$x = c_\ell~(\ell = 1, \ldots, N-1)$ (see~e.g.~\eqref{eq:cpprod}), and only the zero at~$x = c_k$ is encircled by the cycle~$B_k$, we can use~\eqref{eq:simpsqrt} to evaluate~\eqref{eq:adsimp} by residues,\footnote{~\label{ft:cppck}As in footnote~\ref{ft:zeroes}, we evaluate~$C_N^{(0)}(c_k)'' = {(-1)^{k+1} N^2 \over s_k^2}$ by differentiating~\eqref{eq:chebdef}.}
\begin{equation}\label{eq:adfinal}
a_{Dk} = - {i N P_k \over C_N^{(0)}(c_k)'' s_k} = {i (-1)^k\, s_k P_k \over N}~.
\end{equation} 
As expected, $a_{Dk}$ vanishes as~ $P_k \rightarrow 0$. Note that the alternating sign choices for~$P_k$ in \eqref{eq:signs} translate into the statement that all~$a_{Dk} \in i \R_+$.  

For future reference, we substitute~\eqref{eq:adfinal} into~\eqref{eq:lagrange} (and use footnote~\ref{ft:cppck}) to express the polynomial~$P_{N-2}(x)$ in terms of the~$a_{Dk}$,
\begin{equation}\label{eq:pviacp}
P_{N-2}(x) = {i \over N} C_N^{(0)}(x)' \sum_{k = 1}^{N-1} {s_k a_{Dk} \over x- c_k }~.
\end{equation}
We can similarly express~$\delta_k$ in~\eqref{eq:swzeroes} directly in terms of~$a_{Dk}$,
\begin{equation}\label{eq:deltaadrel}
\delta_k^2 = {(-1)^k 2  s_k^2 P_k \over N^2} = -{2 i \over N} s_k a_{Dk}~.
\end{equation}

\section{Expanding the $a$-Periods Around the Multi-Monopole Point} \label{sec:acomp}

\subsection{Setting Up the Computation of the~$a_k$}

In this section we present a direct calculation of the periods~$a_k = {1 \over 2 \pi i} \oint_{A_k} \lambda$ in~\eqref{eq:aaddef} to leading order in small~$P_{N-2}$. As described around equation~\eqref{eq:afromah} our strategy is to calculate the periods~$\hat a_k = {1 \over 2 \pi i} \oint_{\hat A_k} \lambda$, from which the~$a_k$-periods are readily obtained. This calculation is substantially more involved than the calculation of the~$a_D$-periods reviewed in section~\ref{ssec:adreview}. The reason is that~$\hat A_k$ cycles necessarily cross branch cuts as they traverse the two sheets of the Seiberg-Witten curve (see the upper panel of figure~\ref{fig:1}.) Consequently, they cannot be evaluated using residues. A related complication is that the decomposition~$\lambda = \t \lambda + dS$ introduced in~\eqref{eq:lamdecomp} is more delicate, because $\t \lambda$ and~$S$ are not single valued along the~$\hat A_k$ cycles. In particular, the differential~$dS$ (though locally exact) contributes to the integral. 

We begin by converting the period integral over~$\hat A_k$ into an ordinary real integral connecting neighboring branch points of the Seiberg-Witten curve. Taking into account the counterclockwise orientation of the~$\hat A_k$ cycles (see figure~\ref{fig:1}), which leads to a minus sign, and the fact that both the cycles and the Seiberg-Witten differential~$\lambda$ are odd under the hyper-elliptic involution~$(x, y) \rightarrow (x, -y)$, which leads to a factor of~$2$, we can write
\begin{equation}\label{eq:realahint}
\hat a_1 = - {1 \over \pi i} \int_{x_1^+}^{x_0} \lambda~, \qquad \hat a_k = - {1 \over \pi i} \int_{x_k^+}^{x_{k-1}^-} \lambda \qquad (k = 2, \ldots, N-2)~. 
\end{equation}
The locations of the branch points~$x_0, x_k^\pm$ are given by~\eqref{eq:swzeroes} (see also the upper panel of figure~\ref{fig:1}). In the remainder of the paper, we explain how to evaluate the definite integrals in~\eqref{eq:realahint} to leading order in small~$P_{N-2}$. 

Despite the aforementioned subtleties, it is useful to decompose~$\lambda = \t \lambda + dS$ as in~\eqref{eq:lamdecomp}, with~$\t \lambda$ and~$S$ given by~\eqref{eq:ltdef} and~\eqref{eq:sdef}. This leads to a corresponding decomposition of~$\hat a_k$,
\begin{equation}\label{eq:ahdecomp}
\hat a_k = \t a_k + a_k^{(S)}~. 
\end{equation}
Here~$\t a_k$ is the contribution obtained by replacing~$\lambda$ in~\eqref{eq:realahint} with~$\t \lambda$ defined in~\eqref{eq:ltdef},
\begin{equation}\label{eq:atdef}
\begin{split}
 \t a_1 & = - {1 \over \pi i} \int_{x_1^+}^{x_0} \t \lambda \quad = \quad {1 \over \pi i} \int_0^1 d\mu \int_{x_1^+}^{x_0}  dx \, {P_{N-2}(x)  \over \sqrt{\big(C_N^{(\mu)}(x)\big)^2 - 1}}~, \\
\t a_{k \geq 2} & = - {1 \over \pi i} \int_{x_k^+}^{x_{k-1}^-} \t \lambda \; = \; {1 \over \pi i} \int_0^1 d\mu \int_{x_k^+}^{x_{k-1}^-} dx \, {P_{N-2}(x)  \over \sqrt{\big(C_N^{(\mu)}(x)\big)^2 - 1}}~. 
\end{split}
\end{equation}
Analogously, the contribution~$a_k^{(S)}$ in~\eqref{eq:ahdecomp}, which is due to the exact differential~$dS$, reduces to a set of boundary contributions from the limits of the definite integrals in~\eqref{eq:realahint},
\begin{equation}\label{eq:asdef}
 a_1^{(S)} = {1 \over \pi i} \left(S(x_1^+) - S(x_0)\right)~, \qquad a_{k \geq 2}^{(S)} = {1 \over \pi i} \left(S(x_k^+) - S(x_{k-1}^-)\right)~.
\end{equation}

\subsection{The Integrals~$\t a_{k \geq 2}$}\label{ssec:atkgeq2}

We start by evaluating~$\t a_{k \geq 2}$ in~\eqref{eq:atdef},
\begin{equation}\label{eq:takgeq2}
\t a_{k\geq 2} = {1 \over \pi i} \int_0^1 d\mu \int_{x_k^+}^{x_{k-1}^-} dx \, {P_{N-2}(x)  \over \sqrt{\big(C_N^{(\mu)}(x)\big)^2 - 1}}~,
\end{equation}
to leading order in small~$P_{N-2}$. For convenience, we recall some formulas from~\eqref{eq:swzeroes} and~\eqref{eq:cmudef}, 
\begin{equation}
x_k^+ = c_k + \delta_k~, \qquad x_{k-1}^- = c_{k-1} - \delta_{k-1}~, \qquad C_N^{(\mu)}(x) = C_N^{(0)} + \mu P_{N-2}(x)~.
\end{equation}
Although the quantity~$P_{N-2}$ in which we would like to expand appears in the numerator of the integrand, it is not legal to set it to zero in the limits of the integral and under the square root in the denominator. To see this, and to get some intuition for how to proceed, we study the singularities of the integral~\eqref{eq:takgeq2} in more detail.

The polynomial~$\big(C_N^{(\mu)}(x)\big)^2 - 1$ has simple zeroes, which are listed in~\eqref{eq:xmuzeroes}. Together with the product representation for~$\big(C_N^{(0)}(x)\big)^2 - 1$ (implied by~\eqref{eq:ccprel}, \eqref{eq:cpprod}), we deduce
\begin{equation}\label{eq:cnmuprod}
\big(C_N^{(\mu)}(x)\big)^2 - 1 = 2^{2N-2} (x-1+\mu \delta_0)(x+1 - \mu \delta_N) \prod_{k = 1}^{N-1} \left((x-c_k)^2 - \mu \delta_k^2\right)~.
\end{equation} 
This shows that the denominator of the integrand in~\eqref{eq:takgeq2} is non-singular as long as~$0 \leq \mu < 1$. At the endpoint~$\mu = 1$ of the~$\mu$-integral, one simple zero of the polynomial~\eqref{eq:cnmuprod} collides with the endpoint at~$x_k^+$ of the~$x$-integral, while another simple zero of~\eqref{eq:cnmuprod} collides with the other endpoint at~$x_{k-1}^-$. Although the resulting square root singularities are integrable, they modify the expansion of the integral in the small perturbation~$P_{N-2}$.

In order to treat these two singularities, it is convenient to temporarily separate them by introducing a midpoint~$\xi \in (x_k^+, x_{k-1}^-)$ (which is arbitrary but chosen to be well separated from either endpoint, as shown in the upper panel of figure~\ref{fig:1}) and splitting the~$x$-integral in~\eqref{eq:takgeq2} as follows,
\begin{equation}\label{eq:splitxint}
\t a_{k \geq 2} = {1 \over \pi i} \int_0^1 d\mu \int_{x_k^+}^{\xi} dx \, {P_{N-2}(x)  \over \sqrt{\big(C_N^{(\mu)}(x)\big)^2 - 1}}~+~{1 \over \pi i} \int_0^1 d\mu \int_{\xi}^{x_{k-1}^-} dx \, {P_{N-2}(x)  \over \sqrt{\big(C_N^{(\mu)}(x)\big)^2 - 1}}~.
\end{equation}
In the first integral (over~$x \in [x_k^+, \xi ]$) only the singularity at~$x_k^+ = c_k + \delta_k$ is relevant, so that we can set all other~$\delta$'s in~\eqref{eq:cnmuprod} to zero,
\begin{equation}\label{eq:xkpest}
\begin{split}
\big(C_N^{(\mu)}(x)\big)^2 - 1 &~\simeq~2^{2N-2} (x^2-1) \left((x-c_k)^2 - \mu \delta_k^2\right)   \prod_{\substack{\ell = 1\\ \ell \neq k}}^{N-1} (x-c_\ell)^2 \\
&~ =  {(x^2-1) \left((x-c_k)^2 - \mu \delta_k^2\right)  \over N^2 (x-c_k)^2}  \big(C_N^{(0)}(x)'\big)^2~.
\end{split}
\end{equation}
Here we have used~\eqref{eq:cpprod} to obtain the second line. Analogously, only the singularity at~$x_{k-1}^- = c_{k-1} - \delta_{k-1}$ is relevant in the second~$x$-integral (over~$x \in [\xi, x_{k-1}^-]$) in~\eqref{eq:splitxint}, which can therefore be evaluated by approximating
\begin{equation}\label{eq:eq:xkmmest}
\begin{split}
\big(C_N^{(\mu)}(x)\big)^2 - 1 &~\simeq~2^{2N-2} (x^2-1) \left((x-c_{k-1})^2 - \mu \delta_{k-1}^2\right)   \prod_{\substack{\ell = 1\\ \ell \neq k-1}}^{N-1} (x-c_\ell)^2 \\
&~=  {(x^2-1) \left((x-c_{k-1})^2 - \mu \delta_{k-1}^2\right)  \over N^2 (x-c_{k-1})^2}  \big(C_N^{(0)}(x)'\big)^2~.
\end{split}
\end{equation}

We must now take the square roots of~\eqref{eq:xkpest} and~\eqref{eq:eq:xkmmest}, whose sign is fixed by comparing with~\eqref{eq:singsqrt}. Since~$x-c_k>0$ and~$x - c_{k-1} < 0$ have opposite signs on the interval~$[x_k^+, x_{k-1}^-]$, we obtain the following two approximations,
\begin{equation}\label{eq:sqrtapprox}
\begin{split}
\sqrt{\big(C_N^{(\mu)}(x)\big)^2-1}  &~\simeq~ -{i \over N}~{C_N^{(0)}(x)' \over x - c_k} \sqrt{(1-x^2) ((x-c_k)^2-\mu \delta_k^2)}\\
&~\simeq \hskip10pt {i \over N}~{C_N^{(0)}(x)' \over x - c_{k-1}} \sqrt{(1-x^2) ((x-c_{k-1})^2-\mu \delta_{k-1}^2)}~.
\end{split}
\end{equation}
We can use the approximations on the first and second line to simplify the first and second integrals in~\eqref{eq:splitxint}, respectively. Substituting the representation~\eqref{eq:pviacp} for~$P_{N-2}(x)$ into these integrals, we find that the polynomial~$C_N^{(0)}(x)'$ cancels, so that 
\begin{equation}\label{eq:atkg2simp}
\begin{split}
\t a_{k \geq 2} = &~-{1 \over \pi i} \int_0^1 d\mu \int_{c_k + \delta_k}^{\xi} dx \,  {s_k a_{Dk} + (x- c_k) \sum_{\ell = 1, \ell \neq k}^{N-1}  { s_\ell a_{D \ell} \over x- c_\ell} \over \sqrt{(1-x^2) ((x-c_k)^2 - \mu \delta_k^2)} } \\
& + {1 \over \pi i} \int_0^1 d\mu \int_{\xi}^{c_{k-1} - \delta_{k-1}} dx \, {s_{k-1} a_{D,k-1} + (x- c_{k-1}) \sum_{\ell = 1, \ell \neq k-1}^{N-1}  { s_\ell a_{D \ell} \over x- c_\ell} \over \sqrt{(1-x^2) ((x-c_{k-1})^2 - \mu \delta_{k-1}^2)} }~.
\end{split}
\end{equation}
In order to further simplify this integral, we collect terms~$I(a_{Dk})$ whose numerators are proportional to~$a_{Dk}$, terms~$J(a_{D, k-1})$ whose numerators are proportional to~$a_{D, k-1}$, and a remainder~$R(a_{D, \ell \neq k, k -1})$,\footnote{~Below we will see that the integrals~$I(a_{Dk})$, $J(a_{D, k-1})$, and $R(a_{D, \ell \neq k, k -1})$ do in fact only depend on the indicated variables to leading order in small~$a_{D\ell}$. This is no longer the case at higher orders.}
\begin{equation}\label{eq:ijrdef}
\t a_{k \geq 2} \equiv I(a_{Dk}) + J(a_{D, k-1}) + R(a_{D, \ell \neq k, k -1})~.
\end{equation}
We now proceed to define, simplify, and evaluate the integrals~$I(a_{Dk})$, $J(a_{D, k-1})$, and~$R(a_{D, \ell \neq k, k -1})$:

\begin{itemize}
\item[(1.)] The integral~$I(a_{Dk})$ in~\eqref{eq:ijrdef} contains all terms in~\eqref{eq:atkg2simp} whose numerator is proportional to $a_{Dk}$, 
\begin{equation}\label{eq:idef}
\begin{split}
I(a_{Dk}) =~&~-{1 \over \pi i} \int_0^1 d\mu \int_{c_k + \delta_k}^{\xi} dx \,  {s_k a_{Dk} \over \sqrt{(1-x^2) ((x-c_k)^2 - \mu \delta_k^2)} } \\
& +  {1 \over \pi i} \int_0^1 d\mu \int_{\xi}^{c_{k-1} - \delta_{k-1}} dx \, {  s_k a_{D k} \, { x- c_{k-1} \over x- c_k} \over \sqrt{(1-x^2) ((x-c_{k-1})^2 - \mu \delta_{k-1}^2)} }~.
\end{split}
\end{equation}
Since the numerator of the second integral has a simple zero at~$x = c_{k-1}$, and we are only working to leading order in small~$a_{D\ell}$, it is permissible to take~$\delta_{k-1} \rightarrow 0$ in the upper limit of this integral, while approximating its integrand as follows,
\begin{equation}\label{eq:sqrtapproxii}
{ { x- c_{k-1} \over x- c_k} \over \sqrt{(x-c_{k-1})^2 - \mu \delta_{k-1}^2}} \simeq - {1  \over x-c_k} \simeq - {1 \over \sqrt{(x-c_k)^2 - \mu \delta_k^2}}~.
\end{equation}
To leading order in small~$a_{D\ell}$, the two integrals in~\eqref{eq:idef} thus combine into a single integral, which no longer depends on the auxiliary midpoint~$\xi$,
\begin{equation}\label{eq:isimp}
I(a_{Dk}) =~-{1 \over \pi i} \int_0^1 d\mu \int_{c_k + \delta_k}^{c_{k-1}} dx \,  {s_k a_{Dk} \over \sqrt{(1-x^2) ((x-c_k)^2 - \mu \delta_k^2)} }~.
\end{equation}
As explained in appendix~\ref{app:Iint}, this integral can be evaluated explicitly,\footnote{~It is shown in appendix~\ref{app:Iint} that the integrand of~\eqref{eq:isimp} can be expanded in an absolutely convergent power series as long as~$0 \leq \mu < 1$. This power series can then be integrated term-by-term in~$x$ and~$\mu$. Note the crucial role played by the auxiliary parameter~$\mu$ in this approach.} with the following result,
\begin{equation}\label{eq:ifinal}
I(a_{Dk}) =~-{a_{Dk} \over \pi i} \,\left( \log{4 s_k^2 (c_{k-1} - c_k) \over (1 - c_{2k-1}) \delta_k} - {1 \over 2 }\right)~.
\end{equation}

\item[(2.)] The integral~$J(a_{D,k-1})$ in~\eqref{eq:ijrdef} consists of all terms in~\eqref{eq:atkg2simp} whose numerator is proportional to~$a_{D,k-1}$, 
\begin{equation}\label{eq:jdef}
\begin{split}
J(a_{D,k-1}) =~&~-{1 \over \pi i} \int_0^1 d\mu \int_{c_k + \delta_k}^{\xi} dx \,  {s_{k-1} a_{D,k-1} {x- c_k \over x-c_{k-1}} \over \sqrt{(1-x^2) ((x-c_k)^2 - \mu \delta_k^2)} } \\
& +  {1 \over \pi i} \int_0^1 d\mu \int_{\xi}^{c_{k-1} - \delta_{k-1}} dx \, {  s_{k-1} a_{D, k-1}  \over \sqrt{(1-x^2) ((x-c_{k-1})^2 - \mu \delta_{k-1}^2)} }~.
\end{split}
\end{equation}
In exact analogy with the discussion around~\eqref{eq:sqrtapproxii}, we take~$\delta_k \rightarrow 0$ in the first integral and rewrite its integrand so that it can be combined with the second integral. In total, 
\begin{equation}\label{eq:jsimp}
J(a_{D,k-1}) ={1 \over \pi i} \int_0^1 d\mu \int_{c_k}^{c_{k-1} - \delta_{k-1}} dx \, {  s_{k-1} a_{D, k-1}  \over \sqrt{(1-x^2) ((x-c_{k-1})^2 - \mu \delta_{k-1}^2)} }~.
\end{equation}
By comparing with~\eqref{eq:isimp}, we see that the integrals~$J(a_{D, k-1})$ and~$I(a_{D, k})$ are related by a suitable redefinition of parameters. This redefinition is explained in appendix~\ref{app:Jint}, where we show that 
\begin{equation}\label{eq:jfinal}
J(a_{D,k-1}) =~{a_{D,k-1} \over \pi i} \,\left( \log{4 s_{k-1}^2 (c_{k-1} - c_k) \over (1 - c_{2k-1}) \delta_{k-1}} - {1 \over 2 }\right)~.
\end{equation}

\item[(3.)] The remainder~$R(a_{D, \ell \neq k, k -1})$ in~\eqref{eq:ijrdef} consists of all terms in~\eqref{eq:atkg2simp} whose numerators do not contain~$a_{Dk}$ or~$a_{D, k-1}$. Making approximations analogous to those we applied to~$I(a_{Dk})$ and~$J(a_{D, k-1})$ above, we can express
\begin{equation}\label{eq:rdef}
R(a_{D, \ell \neq k, k -1}) = - {1 \over \pi i} \sum_{\substack{\ell = 1 \\ \ell \neq k, k-1}}^{N-1} \int_{c_k}^{c_{k-1}} {dx \over \sqrt{1-x^2}} \, {s_\ell a_{D \ell} \over x- c_\ell }~.
\end{equation}
It is straightforward to evaluate this integral using substitution (see appendix~\ref{app:Rint}),
\begin{equation}\label{eq:rfinal}
R(a_{D, \ell \neq k, k -1}) =  -{1 \over \pi i} \sum_{\substack{\ell = 1 \\ \ell \neq k, k-1}}^{N-1} a_{D \ell} \log{(c_{k-1} - c_\ell) (1-c_{k+\ell}) \over (c_k - c_\ell) (1- c_{k + \ell - 1} ) }~. 
\end{equation}

\end{itemize}

\noindent Finally, we substitute~\eqref{eq:ifinal}, \eqref{eq:jfinal}, \eqref{eq:rfinal} into~\eqref{eq:ijrdef} to obtain a formula for~$\t a_{k \geq 2}$,
\begin{equation}
\begin{split}\label{eq:atkg2final}
\t a_{k \geq 2} =~&  -{a_{Dk} \over \pi i} \,\left( \log{4 s_k^2 (c_{k-1} - c_k) \over (1 - c_{2k-1}) \delta_k} - {1 \over 2 }\right)  + {a_{D,k-1} \over \pi i} \,\left( \log{4 s_{k-1}^2 (c_{k-1} - c_k) \over (1 - c_{2k-1}) \delta_{k-1}} - {1 \over 2 }\right) \\
& -{1 \over \pi i} \sum_{\substack{\ell = 1 \\ \ell \neq k, k-1}}^{N-1} a_{D \ell} \log{(c_{k-1} - c_\ell) (1-c_{k+\ell}) \over (c_k - c_\ell) (1- c_{k + \ell - 1} ) }~. 
\end{split}
\end{equation}

\subsection{The Integral~$\t a_1$}

We now evaluate~$\t a_1$ in~\eqref{eq:atdef},
\begin{equation}\label{eq:at1int}
\t a_1 = {1 \over \pi i} \int_0^1 d\mu \int_{x_1^+}^{x_0}  dx \, {P_{N-2}(x)  \over \sqrt{\big(C_N^{(\mu)}(x)\big)^2 - 1}}~,
\end{equation}
to leading order in small~$P_{N-2}$. Recall from~\eqref{eq:swzeroes}, \eqref{eq:signs} that 
\begin{equation}
x_1^+ = c_1 + \delta_1~, \qquad \delta_1 > 0~, \qquad x_0 = 1- \delta_0~, \qquad \delta_0 = {P_{N-2}(1) \over N^2} <0~, 
\end{equation} 
and from~\eqref{eq:cnmuprod} that~$\big(C_N^{(\mu)}(x)\big)^2 - 1$ has a simple zero at~$x = 1- \mu \delta_0$ that does not collide with any other zeros in the limit~$P_{N-2}(x) \rightarrow 0$. To the order of interest to us, we can therefore evaluate the integral~\eqref{eq:at1int} by approximating the upper limit of the~$x$-integral as~$x_0 \simeq 1$. Following the same logic as for the~$\t a_{k \geq 2}$ in section~\ref{ssec:atkgeq2} above, we can now approximate the square root in the denominator using the first line of~\eqref{eq:sqrtapprox} for~$k = 1$ and all~$x \in [x_1^+, 1]$. As before, we then substitute~\eqref{eq:pviacp} for~$P_{N-2}(x)$ to obtain the following simplification of the integral~\eqref{eq:at1int},
\begin{equation}\label{eq:a1tsimp}
\t a_1 = -{1 \over \pi i} \int_0^1 d\mu \int_{c_1 + \delta_1}^{1}  dx \, {s_1 a_{D1} + \sum_{\ell=2}^{N-1} s_\ell a_{D\ell} {x-c_1 \over x-c_\ell}   \over \sqrt{(1-x^2)((x-c_1)^2 - \mu \delta_1^2)}}~.
\end{equation}
We can set~$\delta_1 \rightarrow 0$ in the second term, so that
\begin{equation}
\label{eq:a1tsimpii}
\begin{split}
\t a_1 =~& -{1 \over \pi i} \int_0^1 d\mu \int_{c_1 + \delta_1}^{1}  dx \, {s_1 a_{D1}  \over \sqrt{(1-x^2)((x-c_1)^2 - \mu \delta_1^2)}} 
- {1 \over \pi i} \int_0^1 d \mu \int_{c_1}^1 {dx \over \sqrt{1-x^2} } \, \sum_{\ell=2}^{N-1} { s_\ell a_{D\ell} \over  x-c_\ell}~.  
\end{split}
\end{equation}
Comparing with~\eqref{eq:isimp}, we see that the first integral in~\eqref{eq:a1tsimpii} is exactly~$I(a_{D 1})$, while comparing with~\eqref{eq:rdef} shows that the second integral in~\eqref{eq:a1tsimpii} is~$R(a_{D, \ell \neq 1})$, i.e.~both can be obtained by setting~$k = 1$ in~\eqref{eq:isimp} and~\eqref{eq:rdef}. Evaluating these integrals by setting~$k =1$ in~\eqref{eq:ifinal} and~\eqref{eq:rfinal}, we find that
\begin{equation}\label{eq:at1final}
\t a_1 = -{a_{D1} \over \pi i} \left(\log\left({ 4 s_1^2 \over \delta_1} \right) - \half\right) - {1 \over \pi i} \sum_{\ell = 2}^{N-1} a_{D \ell} \log{1- c_{\ell+1} \over c_1 - c_\ell}~. 
\end{equation}
Note that this coincides with~\eqref{eq:atkg2final} evaluated at~$k = 1$, as long as we declare that~$a_{D0}=0$.

\subsection{The Boundary Terms~$a_{k\geq 2}^{(S)}$}\label{ssec:askgeq2}

We begin by evaluating the boundary contributions~$a_{k \geq 2}^{(S)}$ in~\eqref{eq:asdef}, 
\begin{equation}\label{eq:asbis}
a_{k \geq 2}^{(S)} = {1 \over \pi i} \left(S(x_k^+) - S(x_{k-1}^-)\right)~.
\end{equation}
The function~$S(x)$ was defined in~\eqref{eq:sdef}, which we repeat here,
\begin{equation}\label{eq:sdefii}
S(x) = x \log \left({C^{(1)}_N(x) + \sqrt{\big(C_N^{(1)}(x)\big)^2 - 1} \over C^{(0)}_N(x) + \sqrt{\big(C_N^{(0)}(x)\big)^2 - 1}}\right) - i N \sqrt{1-x^2}~,
\end{equation} 
where~$C_N^{(1)}(x) = C_N^{(0)}(x) + P_{N-2}(x)$. The location of the branch points~$x_k^\pm$ is given by~\eqref{eq:swzeroes} (see also~\eqref{eq:signs}),
\begin{equation}
x_k^\pm = c_k \pm \delta_k~, \qquad \delta_k > 0~, \qquad (k = 1, \ldots, N-1)~.
\end{equation}
Since the~$x_k^\pm$ are zeros of~$\big(C_N^{(1)}(x)\big)^2 - 1$, it follows that~$C_N^{(1)}(x_k^\pm) = \pm 1$. We can also use~\eqref{eq:chebdef} to show that~$C_N^{(0)}(c_k) = (-1)^k$. Since these expressions must agree as~$P_{N-2} \rightarrow 0$, we obtain the following exact statement,
\begin{equation}\label{eq:c1c0sign}
C_N^{(1)}(x_k^\pm) = C_N^{(0)}(c_k) = (-1)^k~.
\end{equation}
Substituting into~\eqref{eq:sdefii}, we find that 
\begin{equation}\label{eq:sisfk}
S(x_k^\pm) = f_k(x_k^\pm)~,
\end{equation}
where the function~$f_k(x)$ is defined as follows,
\begin{equation}\label{eq:fkdef}
f_k(x) = - x \log \left((-1)^k \left(C^{(0)}_N(x) + \sqrt{\big(C_N^{(0)}(x)\big)^2 - 1} \right)\right) - i N \sqrt{1-x^2}~.
\end{equation}
Here we must use~\eqref{eq:singsqrt} to fix the branch of the square root. It is now straightforward to expand this function around~$x = c_k$,\footnote{~We use~\eqref{eq:chebdef} to compute derivatives of~$C_N^{(0)}(x)$, yielding~$f_k(c_k) = - i N s_k$, $f'_k(c_k) = 0$, and~$f_k''(c_k) = -{i N \over s_k}$.} which in turn gives
\begin{equation}\label{eq:stexp}
S(x_k^\pm) = f_k(c_k \pm \delta_k) = - i N s_k -{i N \over 2 s_k} \delta_k^2 + \cdots~.
\end{equation}
Substituting into~\eqref{eq:asbis} and using~$\delta_k^2 = - {2i \over N} s_k a_{Dk}$ (see~\eqref{eq:deltaadrel}), we obtain
\begin{equation}\label{eq:askgeq2ifn}
a_{k \geq 2}^{(S)} = -{N \over \pi} \left(s_k - s_{k-1}\right) + {i \over \pi} \left(a_{Dk} - a_{D, k-1}\right)~.
\end{equation}

\subsection{The Boundary Term~$a_1^{(S)}$}\label{ssec:a1scomp}

The last contribution we will need is~$a_1^{(S)}$ in~\eqref{eq:asdef},  
\begin{equation}\label{eq:as1}
 a_1^{(S)} = {1 \over \pi i} \left(S(x_1^+) - S(x_0)\right)~.
\end{equation}
In~\eqref{eq:stexp} we have already evaluated  
\begin{equation}\label{eq:sx1p}
S(x_1^+) = - i N s_1 -{i N \over 2 s_1} \delta_1^2~.
\end{equation}
We must now calculate~$S(x_0)$, where~$x_0 = 1 - \delta_0$, with~$\delta_0 = {1 \over N^2} P_{N-2}(1) < 0$, is a simple zero of~$\big(C_N^{(1)}(x)\big)^2 - 1$ (see~\eqref{eq:swzeroes} and~\eqref{eq:signs}). Since it follows from~\eqref{eq:deltaadrel} that~$P_{N-2}(x)$ -- and hence~$\delta_0$ -- is linear in the~$a_{Dk}$, we are free to drop terms beyond first order in~$\delta_0$. In fact, we will show that~$S(x_0)$ vanishes to this order,
\begin{equation}\label{eq:sx0van}
S(x_0) \simeq 0~.
\end{equation}

To see this we must -- for the first and only time in this paper -- explicitly contend with the branch cuts of~$S(x)$. As explained below~\eqref{eq:defmuint}, these branch cuts lie entirely inside the intervals~$(1, +\infty)$, $(x_k^-, x_k^+)$ and~$(-\infty, -1)$,\footnote{~In~\eqref{eq:stexp} we evaluated~$S(x)$ at the branch points~$x = x_k^\pm$, which lie at the boundary of these intervals. Hence~$S(x)$ is single valued there.} but so does the point~$x_0  = 1 - \delta_0 > 1$. If we naively proceed as in section~\ref{ssec:askgeq2} above and attempt to evaluate~$S(x_0)$ by expanding the function~$S(x)$ in~\eqref{eq:sdefii} around~$x = 1$, we find\footnote{~\label{ft:ctaylor}We use~\eqref{eq:c1c0sign} for~$k = 0$ to compute~$C_N^{(1)}(x_0) = C_N^{(0)}(1) = 1$, as well as~\eqref{eq:chebdef} to find~$C_N^{(0)}(1)' = N^2$. In particular, we have~$C_N^{(0)}(1-\delta_0) \simeq 1 - N^2 \delta_0$.}
\begin{equation}\label{eq:sx0naive}
S(1 - \delta_0) \overset{\text{naive}}{\simeq} - N \sqrt{- 2 \delta_0} - i N \sqrt{2 \delta_0}~.
\end{equation}
As we are evaluating~$S(x)$ on one of its branch cuts, it is not surprising that we encounter sign ambiguities. Since~$\delta_0 < 0$, we choose the first square root in~\eqref{eq:sx0naive} to be positive, $\sqrt{- 2 \delta_0} > 0$. As we will see, the relative sign between the two square roots in~\eqref{eq:sx0naive} is then fixed so that the second square root $\sqrt{2 \delta_0} = i \sqrt{- 2 \delta_0}$ exactly cancels the first one, leading to~\eqref{eq:sx0van}. 

Importantly, this cancellation must occur on physical grounds:~the contributions to the~$a$-periods computed in~\eqref{eq:atkg2final} and~\eqref{eq:at1final} already saturate the required monodromies around the multi-monopole point (see~\eqref{eq:akanswerintro}). Therefore all other contributions must be analytic in the~$a_{Dk}$, which would not be the case if the square roots in~\eqref{eq:sx0naive} did not cancel. To see how this cancellation comes about explicitly, we must reexamine the origins of the two square roots in turn. 

The first square root term~$- N \sqrt{- 2 \delta_0} < 0$ in~\eqref{eq:sx0naive} comes from expanding the logarithmic term in~\eqref{eq:sdefii}. As explained in section~\ref{ssec:swdiff}, this term arises from the integral in~\eqref{eq:defmuint}. It follows that the square root that appears in the denominator of that integral must be positive. To see this explicitly, we examine the first integral~$\int_0^1 d\mu\, {x P_{N-2}(x) \over \sqrt{\big(C^{(\mu)}_N(x)\big)^2-1}}$ in~\eqref{eq:defmuint} at~$x = x_0 = 1- \delta_0$. Since~$P_{N-2}(x_0) \simeq P_{N-2}(1) < 0$ (see~\eqref{eq:signs}), it follows that the square root in the denominator of the integral must be positive. Equivalently, we can analyze the second form of the integral~$x \int_{C_N^{(0)}(x)}^{C_N^{(1)}(x)} {d \t \mu \over \sqrt{ \t \mu^2 - 1}}$ in~\eqref{eq:defmuint} at~$x = x_0 = 1- \delta_0$. Since~$C_N^{(1)}(x_0) = 1$ and~$C_N^{(0)}(x_0) \simeq 1 - N^2 \delta_0 > 1$ (see footnote~\ref{ft:ctaylor}), the limits of integration render the integral negative as long as the square root in the denominator is positive. 

The second square root~$-iN \sqrt{2 \delta_0}$ in~\eqref{eq:sx0naive} comes from expanding the pure square root term in~\eqref{eq:sdefii}. As was also explained in section~\ref{ssec:swdiff}, this term ultimately arises from integrating the total $x$-derivative in~\eqref{eq:singswdiff} and picking up the boundary contribution at~$x = x_0$. We can isolate this boundary contribution by integrating~\eqref{eq:singswdiff} from~$x = 1$ to~$x = x_0 = 1- \delta_0 > 1$, since the boundary contribution at~$x = 1$ vanishes,
\begin{equation}\label{eq:signsqrt}
-i \sqrt{2 \delta_0} \simeq - i N \sqrt{1- x_0^2} = \int_1^{x_0} dx \,  {x C^{(0)}_N(x)'  \over \sqrt{\big(C_N^{(0)}(x)\big)^2-1}}~.
\end{equation}
Since~$C_N^{(0)}(x)'$ is positive over the integration region (which can be shown using footnote~\ref{ft:ctaylor}), we conclude that the sign of~\eqref{eq:signsqrt} is set by the sign of the square root in the denominator of the integrand on the right-hand side. However, this sign is not independent. Rather, it must coincide with the signs of the square roots that appear in the integrands of the two integrals in~\eqref{eq:defmuint}. (This ultimately follows from the identity~\eqref{eq:dmuid}, from which all results in section~\ref{ssec:swdiff} follow.) As we explained above, the latter signs must be positive to render the first  term in~\eqref{eq:sx0naive} negative. We thus conclude that the square root in the integrand of~\eqref{eq:signsqrt}, and hence the whole integral, is in fact positive. This completes the proof that the second square root~$-i N \sqrt{2 \delta_0} = N \sqrt{- 2 \delta_0}> 0$ in~\eqref{eq:sx0naive} is positive and cancels the negative first square-root term, which leads to~\eqref{eq:sx0van}.
 
Finally, we can substitute~\eqref{eq:sx1p}, \eqref{eq:sx0van} into~\eqref{eq:as1} and use~$\delta_1^2 = - {2i \over N} s_1 a_{D1}$ (see~\eqref{eq:deltaadrel}) to express our final answer for~$a_1^{(S)}$ as follows,
\begin{equation}\label{eq:a1sfinal}
a_1^{(S)} = -{N \over \pi} s_1 + {i \over  \pi}  a_{D1}~.
\end{equation}
Note that this coincides with~\eqref{eq:askgeq2ifn} evaluated at~$k = 1$, as long as we set~$a_{D0} = 0$.

\subsection{Final Result for the~$a_k$}

We will now combine our preceding results to determine the~$a$-periods via~\eqref{eq:afromah} and~\eqref{eq:ahdecomp},
\begin{equation}\label{eq:acompht}
a_k  = \sum_{\ell = 1}^k \hat a_\ell~, \qquad \hat a_k = \t a_k + a_k^{(S)}~, \qquad (k = 1, \ldots, N-1)~.
\end{equation}
We begin by assembling the answer for~$\hat a_k$. As explained below~\eqref{eq:at1final} and~\eqref{eq:a1sfinal}, we can use~\eqref{eq:atkg2final} for~$\t a_k$ and~\eqref{eq:askgeq2ifn} for~$a_k^{(S)}$ for all~$k = 1, \ldots, N-1$, as long as we set~$a_{D 0} = 0$ in these formulas. Substituting into~\eqref{eq:acompht} and simplifying, we find
\begin{equation}
\begin{split}
\hat a_k ~&= \t a_k + a_k^{(S)} \\
& = -{N \over \pi} \left(s_k - s_{k-1}\right) + {a_{Dk} \over 2 \pi i} \left(\log\left({\delta_k^2 \over 16 s_k^4}\right) -1 \right) - {a_{D, k-1} \over 2 \pi i} \left(\log\left({\delta_{k-1}^2 \over 16 s_{k-1}^4 }\right)-1\right) \\
& \hskip20pt + {1 \over 2 \pi i} \sum_{\substack{\ell=1 \\ \ell \neq k}}^{N-1} a_{D \ell} \log {(c_k - c_\ell)^2 \over (1 - c_{k + \ell})^2}  - {1 \over 2 \pi i} \sum_{\substack{\ell=1 \\ \ell \neq k-1}}^{N-1} a_{D \ell} \log{(c_{k-1} - c_\ell)^2 \over (1- c_{k + \ell -1})^2}~.
\end{split}
\end{equation}
Therefore the sum for~$a_k$ in~\eqref{eq:acompht} telescopes, so that
\begin{equation}
a_k = -{N \over \pi} s_k+ {a_{Dk} \over 2 \pi i} \left(\log\left({\delta_k^2 \over 16 s_k^4}\right) -1 \right) + {1 \over 2 \pi i} \sum_{\substack{\ell=1 \\ \ell \neq k}}^{N-1} a_{D \ell} \log {(c_k - c_\ell)^2 \over (1 - c_{k + \ell})^2}~.
\end{equation}
Finally, we substitute~$\delta_k^2 = - {2 i \over N} s_k a_{Dk}$ from~\eqref{eq:deltaadrel} to obtain the final answer,
\begin{equation}\label{eq:aFinal}
a_k = -{N \over \pi} s_k+ {a_{Dk} \over 2 \pi i} \left(\log\left({-i a_{Dk} \over 8 N s_k^3}\right) -1 \right) + {1 \over 2 \pi i} \sum_{\substack{\ell=1 \\ \ell \neq k}}^{N-1} a_{D \ell} \log {(c_k - c_\ell)^2 \over (1 - c_{k + \ell})^2} ~.
\end{equation}
In order to make contact with the formulas in the introduction, we restore the strong coupling scale~$\Lambda$ by suitably inserting~$1 = 2 \Lambda$ into~\eqref{eq:aFinal},\footnote{~Recall that we set~$\Lambda = \half$ around~\eqref{eq:swdiffnoL}, and that both~$a_k$ and~$a_{Dk}$ have mass-dimension one.} and by using trigonometric identities to simplify the argument of the second logarithm in~\eqref{eq:aFinal}, 
\begin{equation}\label{eq:aFinalLamb}
a_k = -{2 N\Lambda \over \pi} s_k+ {a_{Dk} \over 2 \pi i} \left(\log\left({-i a_{Dk} \over 16 N \Lambda s_k^3}\right) -1 \right) + {1 \over 2 \pi i} \sum_{\substack{\ell=1 \\ \ell \neq k}}^{N-1} a_{D \ell} \log  {{ \sin^2{(k+\ell) \pi \over 2 N } } \over { \sin^2{(k-\ell) \pi \over 2 N } } }~.
\end{equation}

\section*{Acknowledgements}\noindent The work of ED is supported in part by NSF grant PHY-19-14412. TD and EN are supported by a DOE Early Career Award under DE-SC0020421. TD is also supported by a Hellman Fellowship and the Mani L. Bhaumik Presidential Chair in Theoretical Physics at UCLA. The work of EG is in part supported by the Israel National Postdoctoral Award Program for Advancing Women in Science.

\appendix

\section{Comparison with Douglas and Shenker}\label{app:DScomp}

In this appendix we review some of the results obtained in~\cite{Douglas:1995nw} in our conventions. We then extend these results to obtain an alternative derivation of the threshold matrix~$t_{k\ell}$ in~\eqref{eq:tmatansint}. 

\subsection{Review}

First, we show that our Seiberg-Witten curve, as well as our strong-coupling scale~$\Lambda$, are identical to those used in~\cite{Douglas:1995nw}. {By contrast,  our Seiberg-Witten differential~$\lambda$ differs from their differential~$\tilde{\lambda}$ by a sign, i.e.~$\tilde{\lambda} = - {\lambda}$. Since our~$A$- and~$B$-cycles agree with theirs (see footnotes~\ref{ft:dscycle1}--\ref{ft:dscycle3}), this means that their~$\t a_k$- and~$\t a_{Dk}$-periods differ from our~$a_k$- and~$a_{Dk}$-periods by an overall sign, i.e.~$\left(\tilde{a}_{k},\tilde{a}_{Dk}\right) = \left(- {a}_{k},  - a_{Dk}\right)$. Note that these signs cancel in~$\tau_{Dk\ell} = \frac{\partial a_k}{\partial a_{D\ell}} = \t \tau_{D k\ell}$, so that our gauge couplings agree.}

To see this explicitly, let us denote the strong-coupling scale of~\cite{Douglas:1995nw} by~$\t \Lambda$. In units where~$\t \Lambda = 1$, the Seiberg-Witten curve and differential used in~\cite{Douglas:1995nw} take the following form (see the discussion around (2.5) in~\cite{Douglas:1995nw}), 
\begin{equation}\label{eq:dsmatch}
y^2 = \t P(\t x)^2 -1~, \qquad \t P( \t x) = \half  \t x^N + \CO( \t x^{N-2})~, \qquad  \t \lambda =   { \t x d\t  P \over  y}~. 
\end{equation}
We now change variables by writing~$\t x = 2 x$. Comparing with~\eqref{eq:swcurve} and \eqref{eq:noxsq} we see that our conventions for the Seiberg-Witten curve match if we identify~$C_N(x) = \t P(\t x= 2 x)$. Substituting into~\eqref{eq:dsmatch} we see that~$\t \lambda = {2 x C'_N(x) dx /y}$ {appears to} match our Seiberg-Witten differential~$\lambda$ in~\eqref{eq:swdiff} if we set~$\Lambda = 1$, {so that} the strong-coupling scales~$\Lambda$ and~$\t \Lambda$ also agree. {In fact, the two differentials differ by a sign, $\t \lambda = - \lambda$, because the authors of~\cite{Douglas:1995nw} choose the opposite branch of the square root in equation~\eqref{eq:singsqrt} (see their equation~(2.9)), and hence the opposite sign for~$y$. For the remainder of this appendix we work in our conventions, i.e.~we use our Seiberg-Witten differential~$\lambda$, and we set~$\Lambda = \half$ unless otherwise indicated.} 

The scaling trajectory of~\cite{Douglas:1995nw} is given by~$\t P^{(s)}(\t x) = e^s \t P^{(0)}(e^{-s/N} \t x)$, where~$\t P^{(0)}(\t x) = \cos \left( N \arccos {\t x \over 2}\right)$. (See the discussion below equation (5.1) in~\cite{Douglas:1995nw}.) Comparing with~\eqref{eq:chebdef} and the discussion above, we see that the scaling trajectory in our conventions is given by
\begin{equation}\label{eq:cnsdef}
C_N^{(s)}(x) = e^s \t P^{(0)}\left(e^{-{s \over N}} 2 \t x\right) = e^{s} C_N^{(0)}\left(e^{-{s \over N}} x\right)~.
\end{equation}
Here~$C_N^{(0)}(x) = \cos (N \arccos x)$ is the Chebyshev polynomial in~\eqref{eq:chebdef} that describes the singular curve at the multi-monopole point. Expanding~\eqref{eq:cnsdef} to first order in~$s$ and comparing with~\eqref{eq:cc0p}, we find that the degree-$(N-2)$ polynomial describing the approach to the multi-monopole point as~$s\rightarrow 0$ is given by 
\begin{equation}
P_{N-2}(x) = s \left(C_N^{(0)}(x) - {x \over N} C_N^{(0)}(x)'\right) + \CO(s^2)~.
\end{equation}
Note that the leading~$\CO(x^N)$ term cancels out, so that~$P_{N-2}(x)$ does indeed have degree~$N-2$. From this we compute~$P_k = P_{N-2} (c_k) = (-1)^k s$. Substituting into~\eqref{eq:adfinal}, we find that
\begin{equation}\label{eq:addsapp}
a_{Dk} = {i s_k \over N} s + \CO(s^2) = {2 i \Lambda s_k \over N} s + \CO(s^2)~.
\end{equation}
Here we have restored~$\Lambda$, which was previously set to~$\Lambda = \half$. This establishes the formula~\eqref{eq:addsint} quoted in the introduction.\footnote{~Note that setting~$\Lambda = 1$ in~\eqref{eq:addsapp} should reduce to {minus} equation (5.4) in~\cite{Douglas:1995nw}. It does so up to an overall factor {of~$(-2)$ that is missing} in~\cite{Douglas:1995nw}.}   

The authors of~\cite{Douglas:1995nw} compute the magnetic gauge coupling matrix~$\tau_{Dk\ell}(s)$ along their scaling trajectory. They show that this matrix is exactly diagonalizable in a basis of sine functions, so that\footnote{~\label{ft:sinesum}Here we invert equation (5.11) in~\cite{Douglas:1995nw} using~$\sum_{p = 1}^{N-1} s_{p k} s_{p \ell} = {N \over 2} \delta_{k\ell}$.}
\begin{equation}\label{eq:tddsdiag}
\tau_{D k \ell }(s) = {2 \over N} \sum_{p = 1}^{N-1}  \tau_{D}(p, s) s_{p k} s_{p \ell}
\end{equation} 
Here the eigenvalues~$\tau_D(p, s)$ are given in (5.12) of~\cite{Douglas:1995nw} (up to the overall factor of~$i$, which is missing there),
\begin{equation}\label{eq:dsevals}
\tau_D(p, s) = {i \over 2 \sin{\pi p \over 2N}} {F(p, s) \over G(p, s)}~, \qquad (p = 1, \ldots, N-1)~,
\end{equation}
where the functions~$F(p, s)$ and~$G(p,s)$ are defined via the following integrals in (5.9) and (5.10) of~\cite{Douglas:1995nw},
\begin{equation}\label{eq:dsfgdef}
\begin{split}
& F(p, s) = {1 \over \pi} \int_{-b}^b d\theta \, {\cos \left((1-{p \over N}) \theta\right) \over \sqrt{e^{-2s} - \sin^2 \theta}}~, \qquad b = \arcsin e^{-s}~,\\
& G(p, s) = {1 \over \pi} \int_{-a}^a d\theta \, {\cos \left((1-{p \over N}) \theta\right) \over \sqrt{\cos^2 \theta - e^{-2s}}}~, \qquad a = \arccos e^{-s}~.
\end{split}
\end{equation}  
In~\cite{Douglas:1995nw} these integrals were only evaluated for small~$s$ and small~${p \over N}$,
\begin{equation}
F(p, s) = {1 \over \pi} \sin{\pi p \over 2 N} (- \log s) + 1 + \CO\left({p \over N}\right) + \CO(s)~, \qquad G(p, s) = 1 + \CO(s)~. 
\end{equation} 
Substituting into~\eqref{eq:dsevals} gives
\begin{equation}\label{eq:tddssmallk}
\tau_D(p, s) = - {i \over 2 \pi} \log s + {i N \over \pi p} + \CO(1) + \CO(s)~,
\end{equation}
where~$\CO(1)$ refers to the expansion in small~${p \over N}$. This agrees with (5.14) in~\cite{Douglas:1995nw} once the answer there is consistently expanded in small~${p \over N}$ (and again including a missing factor of~$i$). 

To compute~$\tau_{D k \ell}(s)$, we substitute~\eqref{eq:tddssmallk} back into~\eqref{eq:tddsdiag} (and use footnote~\ref{ft:sinesum}),\footnote{~Note that some of the formulas below, e.g.~\eqref{eq:largeNtdmn} or~\eqref{eq:ciresult}, are similar to~\eqref{eq:niceformula} in footnote~\ref{ft:nicefootnote}. However, the latter formula for the off-diagonal elements of~$t_{k\ell}$ is exact, while the former equations are only large-$N$ approximations.}
\begin{equation}\label{eq:largeNtdmn}
\tau_{Dk\ell}(s) = - {i \over 2 \pi} \delta_{k\ell} \log s + {2 \over N} \sum_{p =1}^{N-1} \left({i N \over \pi p} + \CO(1) \right) s_{p k} s_{p \ell}  + \CO(s)~.
\end{equation}
The leading logarithm exactly agrees with the one in (5.16) of~\cite{Douglas:1995nw}. We must now analyze the subleading terms in~\eqref{eq:largeNtdmn}, which approach a finite constant as~$s \rightarrow 0$. Following~\cite{Douglas:1995nw} we show that the sum over~$p$ can be reliably evaluated in the large-$N$ limit. To this end, we let~$\rho = {p \over N}$ and convert the sum over~$p$ to an integral over~$\rho$,
\begin{equation}\label{eq:kappaint}
{2 \over N} \sum_{p =1}^{N-1} \left({i N \over \pi p} + \CO(1) \right) s_{p k} s_{p \ell}  \simeq 2 \int_{1 \over N}^{1 - { 1 \over N}} d \rho \left({i \over \pi \rho} + \CO(1) \right) \sin{\pi k \rho} \sin {\pi \ell \rho}~. 
\end{equation}
We distinguish two cases:
\begin{itemize}
\item[1.)] If either~${k \over N}$ or~${\ell \over N}$ vanish as~$N\rightarrow \infty$ then the corresponding sine functions in the integrand of~\eqref{eq:kappaint} vanish at the lower limit of the integral and can be Taylor expanded there. This cancels the~$1 \over \rho$ pole and renders the integral finite in the large-$N$ limit. A reliable computation of this finite contribution requires knowledge of the~$\CO(1)$ terms in~\eqref{eq:kappaint}.

\item[2.)] If both~$k$ and~$\ell$ are~$\CO(N)$ then both sine functions in~\eqref{eq:kappaint} approach non-zero~$\CO(1)$ constants at~$\rho = {1 \over N}$. The integral is therefore dominated by the~${1 \over \rho}$ divergence there, which can be reliably computed without knowing the~$\CO(1)$ terms in~\eqref{eq:kappaint},
\begin{equation}\label{eq:ciresult}
\begin{split}
{2i \over \pi} \int_{1 \over N}^{1 - { 1 \over N}} {d \rho \over \rho} \,  \sin{\pi k \rho} \sin {\pi \ell \rho} & \simeq - {i \over \pi} \int^{1 \over N} {d \rho \over \rho} \, \Big( \cos\big(\pi \rho (k-\ell)\big)  - \cos\big( \pi \rho (k + \ell)\big) \Big) \\
& \simeq -{i \over \pi} \text{Ci}\left({\pi(k -\ell)\over N}\right) + {i \over \pi} \text{Ci}\left({\pi (k + \ell) \over N}\right)~.
\end{split}
\end{equation}
Here~$\text{Ci}(x) = \int_\infty^x {dt \over t} \cos t$ is the cosine integral function, which is bounded away from $x = 0$, but diverges as~$\text{Ci}(x) = \log x + \CO(1)$ when~$x \rightarrow 0$. Here we have performed the computation for~$k > \ell$; the answer for~$k < \ell$ can be inferred by symmetry, and when~$k = \ell$ the first cosine integral function in~\eqref{eq:ciresult} is replaced by~$\log {1 \over N}$. 

Since~$k$ and~$\ell$ are both~$\CO(N)$ (see above), the second cosine integral function in~\eqref{eq:ciresult} is~$\CO(1)$ in the large-$N$ limit. The only way the first cosine integral function can avoid a similar fate is if~${k - \ell \over N}$ vanishes at large~$N$, so that~$\text{Ci}\left({\pi(k -\ell)\over N}\right) = \log {k -\ell\over N} + \CO(1)$. Substituting back into~\eqref{eq:largeNtdmn} we thus find that 
\begin{equation}\label{eq:finalds}
\tau_{Dk \ell }(s) = - {i \over 2 \pi} \delta_{k\ell} \log s +{i \over 2\pi} \log {N^2 \over (k-\ell)^2 } + \CO(1) + \CO(s)~.
\end{equation}
Here the~$\CO(1)$ terms are~$s$-independent and finite in the large-$N$ limit. The second logarithm in~\eqref{eq:finalds} is only reliable if~${k - \ell \over N} \rightarrow 0$ as~$N \rightarrow \infty$. (As explained above, a special case is~$k = \ell$, where we retain the~$\log N^2$ but omit the factor~$(k-\ell)^2$ in the denominator of the logarithm.) If instead~$k - \ell = \CO(N)$, then this logarithm becomes part of the~$\CO(1)$ terms, which were not computed in~\cite{Douglas:1995nw}.\footnote{~Note that~\eqref{eq:finalds} should agree with equation (5.16) in~\cite{Douglas:1995nw} as long as~$k = \alpha N + \hat k$ and~$\ell = \alpha N + \hat \ell$ with~$\alpha = \CO(1)$ and~${\hat k \over N}, {\hat \ell \over N} \rightarrow 0$ as~$N \rightarrow \infty$. Expanding (5.16) in~\cite{Douglas:1995nw} in this regime yields
$$
\tau_{Dk\ell}(s) = -{i \over 2\pi} \delta_{k\ell} \log s +{i \over 2 \pi} \log \left(\cos^2 \pi \alpha  {(k-\ell)^2 \over N^2}\right) + \CO(1) + \CO(s)~.
$$
This only agrees with~\eqref{eq:finalds} if we flip the sign of the second logarithm and restrict~$\alpha \neq \half$. 
} 
\end{itemize}

\subsection{Some New Results}\label{app:dsnew}

We now explain how to extend the results of~\cite{Douglas:1995nw} reviewed above to exactly compute the constant terms in~$\tau_{Dk\ell}(s)$ at small~$s$. To this end, we must expand the function~$F(p, s)$ in~\eqref{eq:dsfgdef} at small~$s$, but work exactly in~$p$. To this end we expand
\begin{equation}
b = \arcsin {e^{-s}} = {\pi \over 2} - \sqrt{2s} + \CO(s^{3/2})~.
\end{equation}
To get our bearings, we begin by substituting this into~\eqref{eq:dsfgdef} and naively expanding both the limits of the integral and the integrand,
\begin{equation}\label{eq:ftextexp}
\begin{split}
F(p, s) & \simeq {2 \over \pi} \int_0^{{\pi \over 2} - \sqrt{2s}} d \theta \, {\cos \left((1-{p \over N}) \theta\right) \over \sqrt{e^{-2s} - \sin^2 \theta}}\\
& \simeq {2 \over \pi} \int_0^{{\pi \over 2} - \sqrt{2s}} d \theta \, {\cos \Big(\big(1-{p \over N}\big) \theta\Big)} \left({1 \over \cos \theta} + {s \over \cos^3 \theta}  + s^2 \left({3 \over 2 \cos^5 \theta} - {1 \over \cos^3 \theta} \right) + \CO(s^3) \right)
\end{split}
\end{equation}
Note that all cosines in the denominators of the integrand diverge at the upper endpoint of the integral when~$s \rightarrow 0$. Let us estimate this divergence by considering
\begin{equation}\label{eq:divtest}
 \int_0^{{\pi \over 2} - \sqrt{2s}} d \theta \, {\cos \Big(\big(1-{p \over N}\big) \theta\Big) \over \cos^d \theta} = - \int_{\pi \over 2} ^{\sqrt{2s}} d \chi \, {\sin \Big( {\pi p \over 2 N} + \big(1-{p \over N} \big) \chi \Big) \over \sin^d \chi}~,  
\end{equation}
where we have changed variables to~$\chi = { \pi \over 2} - \theta$. Since the divergence arises from the vanishing sine in the denominator as~$s \rightarrow 0$, we can extract the leading divergence by Taylor expanding the integrand around~$\chi = 0$, so that~\eqref{eq:divtest} reduces to
\begin{equation}\label{eq:leadingsdiv}
- \sin {\pi p \over 2 N}  \int^{\sqrt{2s}} { d \chi   \over \chi^d} = \sin {\pi p \over 2 N} \begin{cases}
{1 \over d-1} {1 \over (2s)^{d-1 \over 2} } & \text{if } \quad d \neq 1 \\ - \half \log 2s & \text{if } \quad d = 1
\end{cases}~.
\end{equation}
This shows that all terms in the integrand of~\eqref{eq:ftextexp} that are of the form~$s^n (\cos \theta)^{-2n-1}$ contribute either~$\sim \log s$ (if~$n = 0$) or~$\CO(1)$ (if $n \geq 1$) as~$s \rightarrow 0$, while all other terms are subleading. 

In order to resum all leading terms, we expand the square root in~\eqref{eq:ftextexp} using~\eqref{eq:sqrtexp} from appendix~\ref{app:Iint},
\begin{equation}\label{eq:sqrtexpcos}
{1 \over \sqrt{\cos^2 \theta - 2s + \CO(s^2)}} =  {1 \over \cos \theta} + \sum_{n=1}^\infty {\Gamma(n + \half) \over \Gamma(\half) n! } {(2 s)^n \over \cos^{2n+1} \theta} + \cdots~,
\end{equation} 
where the ellipsis denotes all subleading terms of the form~$s^n  (\cos \theta)^{-k}$ with~$k < 2 n + 1$. Substituting back into~\eqref{eq:ftextexp}, we can carry out the~$\chi$ integral over all~$n \geq 1$ terms in~\eqref{eq:sqrtexpcos} using~\eqref{eq:leadingsdiv},
\begin{equation}
\label{eq:fksintmed}
F(p, s) \simeq {2 \over \pi} \int_0^{{\pi \over 2} - \sqrt{2s}} d \theta \, {\cos \Big(\big(1-{p \over N}\big) \theta\Big) \over \cos \theta} + {2 \over \pi} \sin{\pi p \over 2 N} \sum_{n = 1}^\infty {\Gamma(n + \half) \over \Gamma(\half) \, n! \, 2n}~. 
\end{equation}
The sum over~$n$ can be performed using {\tt Mathematica} and evaluates to~$\log 2$.\footnote{~To see this analytically, we can again use~\eqref{eq:sqrtexp} to express
$$
\sum_{n = 1}^\infty {\Gamma(n + \half) \over \Gamma(\half) n! 2n} = \int_0^1 {dx \over x} \left({1 \over \sqrt{1-x^2}} -1\right) = \log 2~.
$$
To show that the integral indeed evaluates to~$\log 2$, we replace its lower limit by~$\ep > 0$ and take~$\ep \rightarrow 0$ at the end. Using~\eqref{eq:cidef} and~\eqref{eq:ciintfinal}, we evaluate~$\int_\ep^1 {dx \over x \sqrt{1-x^2}} = \log{2 \over \ep} + \CO(\ep)$, while~$-\int_\ep^1 {dx \over x } = \log \ep$. Combining the two integrals and taking~$\ep \rightarrow 0$ we obtain~$\log 2$.}

The remaining integral in~\eqref{eq:fksintmed} must be expanded up to and including~$\CO(1)$ for small~$s$. (Note that evaluating its leading divergence using~\eqref{eq:leadingsdiv} only captures the logarithmically divergent piece of the integral.) This can also be done using {\tt Mathematica},  \ 
\begin{equation}\label{eq:mathmint}
{2 \over \pi} \int_0^{{\pi \over 2} - \sqrt{2s}} d \theta \, {\cos \Big(\big(1-{p \over N}\big) \theta\Big) \over \cos \theta} = {2 \over \pi} \sin{\pi p \over 2N} \left(- \half \log s - {3 \over 2} \log 2 - \gamma - \psi\left({p \over 2 N}\right) -{\pi \over 2} \cot {\pi p \over 2 N}\right)~.
\end{equation}
Here~$\gamma$ is Euler's constant and~$\psi(x)$ is the digamma function. Using Gauss' digamma theorem (see for instance equation (29) on page 19 of~\cite{bateman1953higher}), we can evaluate 
\begin{equation}
\psi\left({p \over 2 N}\right) = - \gamma - 2 \log 2 - \log N - {\pi \over 2} \cot {\pi p \over 2N} + 2 \sum_{q =1}^{N-1} c_{pq} \log  \sin {\pi q \over 2N}~,
\end{equation}
where $c_{pq} = \cos{ \pi pq \over N}$, following the notation of~\eqref{eq:ckskdef}.
Substituting back into~\eqref{eq:mathmint}, we find that~\eqref{eq:fksintmed} simplifies to
\begin{equation}
\label{eq:ffinal}
F(p, s) \simeq {2 \over \pi} \sin{\pi p \over 2N} \left( - \half \log {s \over 8 N^2} - 2 \sum_{q = 1}^{N-1} c_{pq} \log  \sin {\pi q \over 2N}  \right)~.
\end{equation}

We now substitute~\eqref{eq:ffinal} into~\eqref{eq:dsevals} to obtain 
\begin{equation}
\tau_D(p, s) = -{i \over 2 \pi} \log {s \over 8 N^2} - {2i \over \pi} \sum_{q = 1}^{N-1} 
c_{pq} \log  \sin {\pi q \over 2N}  + \CO(s)~.
\end{equation}
Finally we are in a position to substitute this into~\eqref{eq:tddsdiag} and compute~$\tau_{Dk\ell}(s)$. To this end we need the sum in footnote~\ref{ft:sinesum}, as well as the following more complicated sum, 
\begin{equation}
\sum_{p = 1}^{N-1} s_{pk} s_{p\ell} c_{pq} = {N \over 4} \left(\delta_{q, |k-\ell|} - \delta_{q, k+\ell} - \delta_{q, 2N - k -\ell}\right)~, \qquad 1 \leq k, \ell, q \leq  N-1~.
\end{equation}
This leads to
\begin{equation}\label{eq:taudmnsimed}
\tau_{Dk\ell}(s) = - {i \over 2\pi} \delta_{k\ell} \log {s \over 8 N^2}  - {i \over 2 \pi} \sum_{q=1}^{N-1}\left(\delta_{q, |k-\ell|} - \delta_{q, k+\ell} - \delta_{q, 2N - k -\ell}\right) \log \sin^2 {\pi q \over 2N}~.
\end{equation}
The remaining sum over~$q$ evaluates to
\begin{equation}
\sum_{q=1}^{N-1}\left(\delta_{q, |k-\ell |} - \delta_{q, k+\ell} - \delta_{q, 2N - k - \ell}\right) \log \sin^2 {\pi q \over 2N} = \begin{cases}
- \log \sin^2 { \pi k \over N} & \text{if } k = \ell~,\\
\log {\sin^2 {\pi (k-\ell) \over 2N} \over \sin^2 {\pi(k+\ell) \over 2N}} & \text{if } k \neq \ell~.
\end{cases}
\end{equation}
Substituting back into~\eqref{eq:taudmnsimed}, we find perfect agreement with~\eqref{eq:dsslice}, which we repeat here, 
\begin{equation}
\tau_{D k \ell}(s) = {i \over 2 \pi} \begin{cases}
- \log s + \log \left(8 N^2  \sin^2 {\pi k \over N} \right) & \text{if} \quad k = \ell~, \\
\log {{ \sin^2{(k+\ell) \pi \over 2 N } } \over { \sin^2{(k-\ell) \pi \over 2 N } } } & \text{if} \quad k \neq \ell
\end{cases}~.
\end{equation}

\section{Evaluating Some Definite Integrals}

\subsection{Evaluating $R(a_{D, \ell \neq k, k -1})$} \label{app:Rint}

We begin by evaluating the integral~\eqref{eq:rdef}, which we repeat here,
\begin{equation}\label{eq:rdefapp}
R(a_{D, \ell \neq k, k -1}) = - {1 \over \pi i} \sum_{\substack{\ell = 1 \\ \ell \neq k, k-1}}^{N-1} \int_{c_k}^{c_{k-1}} {dx \over \sqrt{1-x^2}} \, {s_\ell a_{D \ell} \over x- c_\ell }~.
\end{equation}
We need the following basic integral,
\begin{equation}\label{eq:cidef}
\CI(a, b; c) = \int_a^b {dx \over \sqrt{1-x^2}} \, {1 \over x-c}~, \qquad -1 < a < b < 1~, \qquad c \in (-1,1)-[a,b]~.
\end{equation}
Let us define the following sign factor,
\begin{equation}\label{eq:sigdef}
\sigma = \begin{cases} +1 \quad  \text{if} \quad c < a \\ -1 \quad \text{if} \quad c > b\end{cases}~.
\end{equation}
By comparing with the integral~\eqref{eq:cidef}, we see that~$\sigma = \text{sign}(\CI)$. We proceed to evaluate this integral using several substitutions:
\begin{itemize}
\item Substituting~$u = {1 \over x-c}$, we find that
\begin{equation}
\CI(a, b; c) = \sigma \int_{1 \over b-c}^{1 \over a-c} {du \over \sqrt{s^2 u^2 - 2 c u -1}}~, \qquad s = \sqrt{1-c^2} > 0~.
\end{equation}

\item Changing variables to~$w = s^2 u - c$, we find that
\begin{equation}\label{eq:wint}
\CI(a, b;c) = {\sigma \over s} \int_{1-bc \over b-c}^{1-ac \over a-c} {d w \over \sqrt{w^2 - 1}}~.
\end{equation}

\item Note that the sign of the integration variable~$w$ in~\eqref{eq:wint} is given by~$\text{sign}(w) = \sigma$. We can thus change variables one more time, to~$w = \sigma \cosh \eta$ with~$\eta > 0$, and evaluate
\begin{equation}\label{eq:coshform}
\CI(a,b; c) = {1 \over s} \left(\cosh^{-1}\left({1-ac \over |a-c|}\right) - \cosh^{-1}\left({1-bc \over |b-c|}\right)\right)~.
\end{equation} 

\end{itemize}

\noindent We can further simplify~\eqref{eq:coshform} by using the fact that~$\cosh^{-1} v = \log(v + \sqrt{v^2-1})$, as long as~$v \geq 1$. Since this is indeed the case for the arguments of the~$\cosh^{-1}$ functions in~\eqref{eq:coshform}, we can finally express the integral in the following form,
\begin{equation}\label{eq:ciintfinal}
\CI(a, b;c) = {1 \over s} \log {(b-c) \Big( 1- ac + s \sqrt{1-a^2} \Big) \over (a-c) \Big(1 - bc + s \sqrt{1-b^2}\Big) }~, \qquad s = \sqrt{1-c^2} > 0~.
\end{equation}
We can now apply this to evaluate $R(a_{D, \ell \neq k, k -1}) = - {1 \over \pi i} \sum_{\ell \neq k, k-1} s_\ell a_{D \ell} \CI(c_k, c_{k-1}; c_\ell)$ in~\eqref{eq:rdefapp}, for which we need
\begin{equation}
\CI(c_k, c_{k-1}; c_\ell) = {1 \over s_\ell} \log{(c_{k-1} - c_\ell) (1-c_{k+\ell}) \over (c_k - c_\ell) (1- c_{k + \ell - 1} ) }~, \qquad \ell \neq k, k-1~.
\end{equation}
Here we have used the addition formula~$c_k c_\ell - s_k s_\ell = c_{k + \ell}$ for cosines. Substituting into~\eqref{eq:rdefapp}, we obtain
\begin{equation}\label{eq:rfinalapp}
R(a_{D, \ell \neq k, k -1}) =  -{1 \over \pi i} \sum_{\substack{\ell = 1 \\ \ell \neq k, k-1}}^{N-1} a_{D \ell} \log{(c_{k-1} - c_\ell) (1-c_{k+\ell}) \over (c_k - c_\ell) (1- c_{k + \ell - 1} ) }~. 
\end{equation}

\subsection{Evaluating $I(a_{Dk})$}\label{app:Iint}

We now compute the integral~$I(a_{Dk})$ in~\eqref{eq:isimp}, 
\begin{equation}\label{eq:iapp}
I(a_{Dk}) =~-{s_k a_{Dk} \over \pi i} \, \hat I(a_{Dk})~,
\end{equation}
where the integral~$\hat I (a_{Dk})$ that we must evaluate is given by
\begin{equation}\label{eq:ihatdef}
\hat I(a_{Dk}) = \int_0^1 d\mu \int_{c_k + \delta_k}^{c_{k-1}} dx \,  {1 \over \sqrt{(1-x^2) ((x-c_k)^2 - \mu \delta_k^2)} }~.
\end{equation}
Note that this integral is manifestly positive. We will not retain terms in~$I(a_{Dk})$ that vanish faster than~$a_{Dk}$. For this reason, we can drop terms in~$\hat I(a_{Dk})$ that vanish when~$a_{Dk} \rightarrow 0$, or equivalently when~$\delta_k \rightarrow 0$ (see~\eqref{eq:deltaadrel}). 

We will directly evaluate the integral~\eqref{eq:ihatdef} by expanding both inverse square roots in absolutely convergent power series and integrating term by term.\footnote{~The expansion of the first square root is absolutely convergent in the entire integration region, while the expansion of the second square root is absolutely convergent as long as~$\mu < 1$.} To this end, we expand the first inverse square root via
\begin{equation}\label{eq:sqrtexp}
{1 \over \sqrt{1-x^2}} = \sum_{n=0}^\infty {\Gamma(n+ \half ) \over \Gamma(\half) n!} \, x^{2n}~,
\end{equation}
and similarly for the second inverse square root. After substituting into~\eqref{eq:ihatdef}, we can carry out the~$\mu$ integral. We then simplify the~$x$ integral by shifting~$x \rightarrow x + c_k$ and expanding the numerator using the binomial formula, so that
\begin{equation}\label{eq:ihatintstep}
\hat I(a_{Dk}) = \sum_{m, n = 0}^\infty {\Gamma(m + \half) \Gamma(n + \half) \over \Gamma(\half)^2 m! (n+1)!} \, \delta_k^{2n} \sum_{\ell = 0}^{2m} \begin{pmatrix} 2m \\ \ell \end{pmatrix} c_k^\ell \int_{\delta_k}^{c_{k-1} - c_k} dx \, x^{2m - \ell - 2n - 1}~.
\end{equation}
The remaining~$x$-integral is trivial,
\begin{equation}\label{eq:xint}
\int_{\delta_k}^{c_{k-1} - c_k} dx \, x^{2m - \ell - 2n - 1} = \begin{cases} \log{c_{k-1} - c_k \over \delta_k}  \quad \text{if} \quad 2m - \ell - 2n = 0 \\
{1 \over 2m  - \ell - 2n } \left((c_{k-1} - c_k)^{2m - \ell - 2n} - \delta_k^{2m - \ell - 2n} \right)\quad \text{if} \quad 2m - \ell - 2n \neq 0
 \end{cases}~.
\end{equation}
Substituting back into~\eqref{eq:ihatintstep}, we now drop all terms that vanish as~$\delta_k \rightarrow 0$. The only remaining terms are the~$n = 0$ logarithmic terms and the~$n = 0$ polynomial terms from the upper limit of the $x$-integral~\eqref{eq:xint}, as well as the~$\ell = 2m$ polynomial terms from the lower limit of the same integral. Paying attention to the restrictions on summation indices that result from~\eqref{eq:xint}, we can now express~\eqref{eq:ihatintstep} as a sum of three terms, 
\begin{equation}\label{eq:ihisf}
\hat I(a_{Dk}) = f_1 + f_2 + f_3~,
\end{equation}
where~$f_{1,2,3}$ are given by the following series expressions,
\begin{equation}
\begin{split}\label{eq:fdef}
& f_1  = \log{c_{k-1} - c_k \over \delta_k}  \sum_{m = 0}^\infty {\Gamma(m + \half)  \over \Gamma(\half) m!} \, c_k^{2m}~,\\
& f_2 = \sum_{m = 1}^\infty {\Gamma(m + \half)  \over \Gamma(\half) m!} \, \sum_{\ell = 0}^{2m - 1} \begin{pmatrix} 2m \\ \ell \end{pmatrix} { c_k^\ell (c_{k-1} - c_k)^{2m - \ell} \over 2m - \ell}~,\\
& f_3 = \sum_{m = 0}^\infty \sum_{n = 1}^\infty {\Gamma(m + \half) \Gamma(n + \half) \over \Gamma(\half)^2 m! n! } \, { c_k^{2m} \over 2n (n+1)}~. 
\end{split}
\end{equation}
The sums over~$m$ in~$f_1$ and~$f_3$ can be evaluated using~\eqref{eq:sqrtexp}, while the remaining sum in~$f_3$ can be performed using {\tt Mathematica}. This gives
\begin{equation}\label{eq:f1f3res}
f_1 = {1 \over s_k} \log{c_{k-1} - c_k \over \delta_k} ~, \qquad f_3 = {1 \over s_k} \left(\log 2 - \half\right)~.
\end{equation} 

To evaluate~$f_2$ in~\eqref{eq:fdef}, we define the function~$f_2(x)$ via
\begin{equation}\label{eq:f2xdef}
f_2(x) = \sum_{m = 1}^\infty {\Gamma(m + \half)  \over \Gamma(\half) m!} \, \sum_{\ell = 0}^{2m - 1} \begin{pmatrix} 2m \\ \ell \end{pmatrix} { c_k^\ell (x - c_k)^{2m - \ell} \over 2m - \ell}~,
\end{equation} 
so that
\begin{equation}\label{eq:f2bc}
f_2(c_{k-1}) = f_2~, \qquad f_2(c_k) = 0~.
\end{equation}
Differentiating~\eqref{eq:f2xdef} term by term and summing the resulting series using~\eqref{eq:sqrtexp}, we find
\begin{equation}\label{eq:f2p}
f_2'(x) = {1 \over \sqrt{1-x^2} \, (x-c_k)} - {1 \over s_k(x-c_k)}~.
\end{equation} 
We now integrate this equation from~$a$ to~$x$, where~$c_k < a, x < c_{k-1}$. The first term on the right-hand side leads to an integral of the form~\eqref{eq:cidef}, while the second term integrates to a logarithm,
\begin{equation}
f_2(x) = f_2(a) + \CI(a, x; c_k) - {1 \over s_k} \log{x - c_k \over a - c_k}~.
\end{equation}
If we evaluate~$\CI(a,x; c_k)$ using~\eqref{eq:ciintfinal} and fix the integration constant~$f_2(a)$ by imposing the boundary condition~$f_2(c_k) = 0$ in~\eqref{eq:f2bc}, we find that
\begin{equation}\label{eq:f2xfin}
f_2(x) = {1 \over s_k} \log{2 s_k^2 \over 1- x c_k + s_k \sqrt{1-x^2} }~.
\end{equation} 
Note that the arguments of the square root and the logarithm in this formula are strictly positive for~$c_{k-1} \leq x \leq c_k$, so that~$f_2(x)$ is indeed real analytic on that interval. We can now use~\eqref{eq:f2bc} and~\eqref{eq:f2xfin} to evaluate the second sum~$f_2$ in~\eqref{eq:fdef},
\begin{equation}\label{eq:f2ans}
f_2 = f_2(c_{k-1}) = {1 \over s_k} \log{2 s_k^2 \over 1 - c_{2k-1}}~.
\end{equation}
Here we have used the cosine addition formula~$c_{k} c_{k-1} - s_k s_{k-1} = c_{2k-1}$. 

We are now ready to assemble the answer: substituting~$f_{1, 3}$ in~\eqref{eq:f1f3res} and~$f_2$ in~\eqref{eq:f2ans} into~\eqref{eq:ihisf}, we find that
\begin{equation}\label{eq:ihatfinal}
\hat I(a_{Dk}) = {1 \over s_k} \log{4 s_k^2 (c_{k-1} - c_k) \over (1 - c_{2k-1}) \delta_k} - {1 \over 2 s_k}~.
\end{equation}
As expected (see the comment below~\eqref{eq:ihatdef}), this expression is positive in the limit~$\delta_k \rightarrow 0$, where~$\hat I(a_{D k}) \simeq -{1 \over s_k} \log \delta_k > 0$. Finally, the original integral~\eqref{eq:iapp} evaluates to
\begin{equation}\label{eq:ifinalapp}
I(a_{Dk}) =~-{a_{Dk} \over \pi i} \,\left( \log{4 s_k^2 (c_{k-1} - c_k) \over (1 - c_{2k-1}) \delta_k} - {1 \over 2 }\right)~.
\end{equation}

\subsection{Evaluating $J(a_{D,k-1})$}\label{app:Jint}

Here we evaluate  the integral~$J(a_{D,k-1})$ in~\eqref{eq:jsimp}, 
\begin{equation}\label{eq:japp}
J(a_{D,k-1}) =~ {s_{k-1} a_{D,k-1} \over \pi i} \, \hat J(a_{D,k-1})~,
\end{equation}
where the integral~$\hat J(a_{D,k-1})$ that we must evaluate is given by
\begin{equation}\label{eq:jhatdef}
\hat J(a_{D,k-1}) = \int_0^1 d\mu \int_{c_k}^{c_{k-1}-\delta_{k-1}} dx \,  {1 \over \sqrt{(1-x^2) ((x-c_{k-1})^2 - \mu \delta_{k-1}^2)} }~.
\end{equation}
Note that this integral is manifestly positive. Comparing with~\eqref{eq:ihatdef} makes it clear that it should be possible to evaluate~$\hat J(a_{D,k-1})$ by carefully continuing the parameters that enter the definition of~$\hat I(a_{Dk})$.\footnote{~\label{ft:negdel}A naive continuation that gives wrong answers is~$\hat J(a_{D,k-1}) = - \hat I(a_{Dk})\big|_{c_k \leftrightarrow c_{k-1}, \delta_k \rightarrow - \delta_{k-1}}$. (One way to see that this cannot be correct is that the two sides have opposite signs.) This continuation fails because flipping the sign of~$\delta$ extends the~$x$-integral past a branch point of the square root in the denominator.} We initially proceed as in appendix~\ref{app:Iint}, and derive for~$\hat J(a_{D,k-1})$ the same series representation that we obtained for~$\hat I(a_{Dk})$ in~\eqref{eq:ihatintstep},
\begin{equation}\label{eq:jhatintstep}
\hat J(a_{D,k-1}) = \sum_{m, n = 0}^\infty {\Gamma(m + \half) \Gamma(n + \half) \over \Gamma(\half)^2 m! (n+1)!} \, \delta_{k-1}^{2n} \sum_{\ell = 0}^{2m} \begin{pmatrix} 2m \\ \ell \end{pmatrix} (-c_{k-1})^\ell \int_{\delta_{k-1}}^{c_{k-1} - c_k} dx \, x^{2m - \ell - 2n - 1}~.
\end{equation}
Comparing this with~\eqref{eq:ihatintstep}, we see that we can compute~$\hat J(a_{D,k-1})$ from~$\hat I(a_{Dk})$ by substituting~$c_k \rightarrow - c_{k-1}$, $c_{k-1} \rightarrow -c_k$, and~$\delta_k \rightarrow \delta_{k-1}$.\footnote{~Note that these continuations do not run afoul of the same problems as the ones in footnote~\ref{ft:negdel}.} Substituting these replacements into~\eqref{eq:ihatfinal}, we find that\footnote{~Note that~$c_{k-1} - c_k$, as well as~$s_k = \sqrt{1-c_k^2}$, $s_{k-1} = \sqrt{1-c_{k-1}^2}$, and~$c_{2k -1} = c_k c_{k-1} - s_k s_{k-1}$ are invariant under the substitutions~$c_k \rightarrow - c_{k-1}$, $c_{k-1} \rightarrow -c_k$.}
\begin{equation}\label{eq:jhatfinal}
\hat J(a_{D,k-1}) = \hat I(a_{Dk})\bigg|_{\substack{c_k \rightarrow - c_{k-1} \\ c_{k-1} \rightarrow -c_k\\ \delta_k \rightarrow \delta_{k-1}}} = {1 \over s_{k-1}} \log{4 s_{k-1}^2 (c_{k-1} - c_k) \over (1- c_{2k -1}) \delta_{k-1} } - {1 \over 2 s_{k-1}}~.
\end{equation}
Note that this is positive in the limit~$\delta_{k-1} \rightarrow 0$, where~$\hat J(a_{D,k-1}) \simeq - {1 \over s_{k-1}} \log \delta_{k-1} > 0$, in agreement with the comment below~\eqref{eq:jhatdef}. Substituting~\eqref{eq:jhatfinal} into~\eqref{eq:japp}, we finally obtain
\begin{equation}\label{eq:jappfinal}
J(a_{D, k-1}) = {a_{D, k-1} \over \pi i} \left( \log{4 s_{k-1}^2 (c_{k-1} - c_k) \over (1- c_{2k -1}) \delta_{k-1} } - {1 \over 2} \right)~.
\end{equation}


\bibliographystyle{utphys}
\bibliography{tmatrix}

\providecommand{\href}[2]{#2}\begingroup\raggedright\begin{thebibliography}{10}

\bibitem{Seiberg:1994rs}
N.~Seiberg and E.~Witten, ``{Electric - magnetic duality, monopole
  condensation, and confinement in N=2 supersymmetric Yang-Mills theory},''
  \href{http://dx.doi.org/10.1016/0550-3213(94)90124-4,
  10.1016/0550-3213(94)00449-8}{{\em Nucl. Phys.} {\bfseries B426} (1994)
  19--52}, \href{http://arxiv.org/abs/hep-th/9407087}{{\ttfamily
  arXiv:hep-th/9407087 [hep-th]}}.
[Erratum: Nucl. Phys.B430,485(1994)].

\bibitem{Seiberg:1994aj}
N.~Seiberg and E.~Witten, ``{Monopoles, duality and chiral symmetry breaking in
  N=2 supersymmetric QCD},''
  \href{http://dx.doi.org/10.1016/0550-3213(94)90214-3}{{\em Nucl. Phys.}
  {\bfseries B431} (1994) 484--550},
\href{http://arxiv.org/abs/hep-th/9408099}{{\ttfamily arXiv:hep-th/9408099
  [hep-th]}}.

\bibitem{Witten:1994cg}
E.~Witten, ``{Monopoles and four manifolds},''
  \href{http://dx.doi.org/10.4310/MRL.1994.v1.n6.a13}{{\em Math. Res. Lett.}
  {\bfseries 1} (1994) 769--796},
  \href{http://arxiv.org/abs/hep-th/9411102}{{\ttfamily arXiv:hep-th/9411102}}.

\bibitem{Douglas:1995nw}
M.~R. Douglas and S.~H. Shenker, ``{Dynamics of SU(N) supersymmetric gauge
  theory},'' \href{http://dx.doi.org/10.1016/0550-3213(95)00258-T}{{\em Nucl.
  Phys.} {\bfseries B447} (1995) 271--296},
\href{http://arxiv.org/abs/hep-th/9503163}{{\ttfamily arXiv:hep-th/9503163
  [hep-th]}}.

\bibitem{Klemm:1994qs}
A.~Klemm, W.~Lerche, S.~Yankielowicz, and S.~Theisen, ``{Simple singularities
  and N=2 supersymmetric Yang-Mills theory},''
  \href{http://dx.doi.org/10.1016/0370-2693(94)01516-F}{{\em Phys. Lett.}
  {\bfseries B344} (1995) 169--175},
\href{http://arxiv.org/abs/hep-th/9411048}{{\ttfamily arXiv:hep-th/9411048
  [hep-th]}}.

\bibitem{Argyres:1994xh}
P.~C. Argyres and A.~E. Faraggi, ``{The vacuum structure and spectrum of N=2
  supersymmetric SU(n) gauge theory},''
  \href{http://dx.doi.org/10.1103/PhysRevLett.74.3931}{{\em Phys. Rev. Lett.}
  {\bfseries 74} (1995) 3931--3934},
\href{http://arxiv.org/abs/hep-th/9411057}{{\ttfamily arXiv:hep-th/9411057
  [hep-th]}}.

\bibitem{Klemm:1994qj}
A.~Klemm, W.~Lerche, S.~Yankielowicz, and S.~Theisen, ``{On the monodromies of
  N=2 supersymmetric Yang-Mills theory},'' in {\em {Joint U.S.-Polish Workshop
  on Physics from Planck Scale to Electro-Weak Scale (SUSY 94)}}, pp.~433--446.
\newblock 1995.
\newblock \href{http://arxiv.org/abs/hep-th/9412158}{{\ttfamily
  arXiv:hep-th/9412158}}.

\bibitem{Klemm:1995wp}
A.~Klemm, W.~Lerche, and S.~Theisen, ``{Nonperturbative effective actions of
  N=2 supersymmetric gauge theories},''
  \href{http://dx.doi.org/10.1142/S0217751X96001000}{{\em Int. J. Mod. Phys. A}
  {\bfseries 11} (1996) 1929--1974},
  \href{http://arxiv.org/abs/hep-th/9505150}{{\ttfamily arXiv:hep-th/9505150}}.

\bibitem{Argyres:1995jj}
P.~C. Argyres and M.~R. Douglas, ``{New phenomena in SU(3) supersymmetric gauge
  theory},'' \href{http://dx.doi.org/10.1016/0550-3213(95)00281-V}{{\em Nucl.
  Phys. B} {\bfseries 448} (1995) 93--126},
  \href{http://arxiv.org/abs/hep-th/9505062}{{\ttfamily arXiv:hep-th/9505062}}.

\bibitem{Argyres:1995xn}
P.~C. Argyres, M.~Plesser, N.~Seiberg, and E.~Witten, ``{New N=2 superconformal
  field theories in four-dimensions},''
  \href{http://dx.doi.org/10.1016/0550-3213(95)00671-0}{{\em Nucl. Phys. B}
  {\bfseries 461} (1996) 71--84},
  \href{http://arxiv.org/abs/hep-th/9511154}{{\ttfamily arXiv:hep-th/9511154}}.

\bibitem{Matone:1995rx}
M.~Matone, ``{Instantons and recursion relations in N=2 SUSY gauge theory},''
  \href{http://dx.doi.org/10.1016/0370-2693(95)00920-G}{{\em Phys. Lett. B}
  {\bfseries 357} (1995) 342--348},
  \href{http://arxiv.org/abs/hep-th/9506102}{{\ttfamily arXiv:hep-th/9506102}}.

\bibitem{Eguchi:1995jh}
T.~Eguchi and S.-K. Yang, ``{Prepotentials of N=2 supersymmetric gauge theories
  and soliton equations},''
  \href{http://dx.doi.org/10.1142/S0217732396000151}{{\em Mod. Phys. Lett. A}
  {\bfseries 11} (1996) 131--138},
  \href{http://arxiv.org/abs/hep-th/9510183}{{\ttfamily arXiv:hep-th/9510183}}.

\bibitem{Bonelli:1996qc}
G.~Bonelli and M.~Matone, ``{Nonperturbative renormalization group equation and
  beta function in N=2 supersymmetric Yang-Mills theory},''
  \href{http://dx.doi.org/10.1103/PhysRevLett.76.4107}{{\em Phys. Rev. Lett.}
  {\bfseries 76} (1996) 4107--4110},
  \href{http://arxiv.org/abs/hep-th/9602174}{{\ttfamily arXiv:hep-th/9602174}}.

\bibitem{Bonelli:1996qh}
G.~Bonelli and M.~Matone, ``{Nonperturbative relations in N=2 supersymmetric
  Yang-Mills theory and the Witten-Dijkgraaf-Verlinde-Verlinde equation},''
  \href{http://dx.doi.org/10.1103/PhysRevLett.77.4712}{{\em Phys. Rev. Lett.}
  {\bfseries 77} (1996) 4712--4715},
  \href{http://arxiv.org/abs/hep-th/9605090}{{\ttfamily arXiv:hep-th/9605090}}.

\bibitem{Howe:1996pw}
P.~S. Howe and P.~C. West, ``{Superconformal ward identities and N=2 Yang-Mills
  theory},'' \href{http://dx.doi.org/10.1016/S0550-3213(96)00628-1}{{\em Nucl.
  Phys. B} {\bfseries 486} (1997) 425--442},
  \href{http://arxiv.org/abs/hep-th/9607239}{{\ttfamily arXiv:hep-th/9607239}}.

\bibitem{DHoker:1996yyu}
E.~D'Hoker, I.~Krichever, and D.~Phong, ``{The Renormalization group equation
  in N=2 supersymmetric gauge theories},''
  \href{http://dx.doi.org/10.1016/S0550-3213(97)00156-9}{{\em Nucl. Phys. B}
  {\bfseries 494} (1997) 89--104},
  \href{http://arxiv.org/abs/hep-th/9610156}{{\ttfamily arXiv:hep-th/9610156}}.

\bibitem{Luty:1999qc}
M.~A. Luty and R.~Rattazzi, ``{Soft supersymmetry breaking in deformed moduli
  spaces, conformal theories, and N=2 Yang-Mills theory},''
  \href{http://dx.doi.org/10.1088/1126-6708/1999/11/001}{{\em JHEP} {\bfseries
  11} (1999) 001},
\href{http://arxiv.org/abs/hep-th/9908085}{{\ttfamily arXiv:hep-th/9908085
  [hep-th]}}.

\bibitem{FutureUs}
E.~D'Hoker, T.~T. Dumitrescu, E.~Gerchkovitz, and E.~Nardoni, ``To appear,''.

\bibitem{Cordova:2018acb}
C.~Cordova and T.~T. Dumitrescu, ``{Candidate Phases for SU(2) Adjoint QCD$_4$
  with Two Flavors from $\mathcal{N}=2$ Supersymmetric Yang-Mills Theory},''
\href{http://arxiv.org/abs/1806.09592}{{\ttfamily arXiv:1806.09592 [hep-th]}}.

\bibitem{DHoker:1997mlo}
E.~D'Hoker and D.~H. Phong, ``{Strong coupling expansions of SU(N)
  Seiberg-Witten theory},''
  \href{http://dx.doi.org/10.1016/S0370-2693(97)00145-7}{{\em Phys. Lett.}
  {\bfseries B397} (1997) 94--103},
\href{http://arxiv.org/abs/hep-th/9701055}{{\ttfamily arXiv:hep-th/9701055
  [hep-th]}}.

\bibitem{Edelstein:1999fz}
J.~D. Edelstein and J.~Mas, ``{Strong coupling expansion and
  Seiberg-Witten-Whitham equations},''
  \href{http://dx.doi.org/10.1016/S0370-2693(99)00262-2}{{\em Phys. Lett. B}
  {\bfseries 452} (1999) 69--75},
  \href{http://arxiv.org/abs/hep-th/9901006}{{\ttfamily arXiv:hep-th/9901006}}.

\bibitem{Edelstein:1999tb}
J.~D. Edelstein and J.~Mas, ``{N=2 supersymmetric Yang-Mills theories and
  Whitham integrable hierarchies},''
  \href{http://dx.doi.org/10.1063/1.59658}{{\em AIP Conf. Proc.} {\bfseries
  484} no.~1, (1999) 195--212},
  \href{http://arxiv.org/abs/hep-th/9902161}{{\ttfamily arXiv:hep-th/9902161}}.

\bibitem{Edelstein:2000aj}
J.~D. Edelstein, M.~Gomez-Reino, and M.~Marino, ``{Blowup formulae in
  Donaldson-Witten theory and integrable hierarchies},''
  \href{http://dx.doi.org/10.4310/ATMP.2000.v4.n3.a1}{{\em Adv. Theor. Math.
  Phys.} {\bfseries 4} (2000) 503--543},
  \href{http://arxiv.org/abs/hep-th/0006113}{{\ttfamily arXiv:hep-th/0006113}}.

\bibitem{Braden:2000he}
H.~Braden and A.~Marshakov, ``{Singular phases of Seiberg-Witten integrable
  systems: Weak and strong coupling},''
  \href{http://dx.doi.org/10.1016/S0550-3213(00)00683-0}{{\em Nucl. Phys. B}
  {\bfseries 595} (2001) 417--466},
  \href{http://arxiv.org/abs/hep-th/0009060}{{\ttfamily arXiv:hep-th/0009060}}.

\bibitem{Bonelli:2017ptp}
G.~Bonelli, A.~Grassi, and A.~Tanzini, ``{New results in $\mathcal{N}=2$
  theories from non-perturbative string},''
  \href{http://dx.doi.org/10.1007/s00023-017-0643-5}{{\em Annales Henri
  Poincare} {\bfseries 19} no.~3, (2018) 743--774},
  \href{http://arxiv.org/abs/1704.01517}{{\ttfamily arXiv:1704.01517
  [hep-th]}}.

\bibitem{Gaiotto:2008cd}
D.~Gaiotto, G.~W. Moore, and A.~Neitzke, ``{Four-dimensional wall-crossing via
  three-dimensional field theory},''
  \href{http://dx.doi.org/10.1007/s00220-010-1071-2}{{\em Commun. Math. Phys.}
  {\bfseries 299} (2010) 163--224},
  \href{http://arxiv.org/abs/0807.4723}{{\ttfamily arXiv:0807.4723 [hep-th]}}.

\bibitem{bateman1953higher}
A.~Erdelyi, {\em Higher transcendental functions, Volume 1 (Bateman Manuscript
  project)}.
\newblock Krieger, 1981.

\end{thebibliography}\endgroup

\end{document}